\documentclass[fleqn,usenatbib]{mnras}

\usepackage{newtxtext,newtxmath}

\usepackage[T1]{fontenc}
\usepackage{ae,aecompl}

\usepackage{graphicx}	
\usepackage{amsmath}	

\usepackage{multirow}
\usepackage{adjustbox}

\usepackage[verbose]{placeins}
\usepackage{blindtext}

\usepackage{bm}
\usepackage{mathtools}
\usepackage{xcolor}
\usepackage{natbib}
\usepackage{lscape}
\usepackage{afterpage}  
\usepackage{booktabs}
\usepackage{tabularx}
\usepackage{threeparttable}

\usepackage{newtxtext,newtxmath}

\title[A strong LyC leaker at $z=3.613$]{An extreme blue nugget, UV-bright starburst at $z=3.613$ with ninety per cent of Lyman continuum photon escape}

\author[R. Marques-Chaves et al.]{
R. Marques-Chaves$^{1}$\thanks{E-mail: Rui.MarquesCoelhoChaves@unige.ch},
D. Schaerer$^{1,2}$, 
J. \'{A}lvarez-M\'{a}rquez$^{3}$, 
A. Verhamme$^{1}$, 
D. Ceverino$^{4,5}$,
\newauthor 
J. Chisholm$^{6}$,
L. Colina$^{3,7}$, 
M. Dessauges-Zavadsky$^{1}$,
I. P\'{e}rez-Fournon$^{8,9}$,
A. Saldana-Lopez$^{1}$,
\newauthor 
A. Upadhyaya$^{1}$,
E. Vanzella$^{10}$\\
\\
$^{1}$Geneva Observatory, Department of Astronomy, University of Geneva, Chemin Pegasi 51, CH-1290 Versoix, Switzerland\\
$^{2}$CNRS, IRAP, 14 Avenue E. Belin, 31400 Toulouse, France\\
$^{3}$Centro de Astrobiolog\'ia (CSIC-INTA), Carretera de Ajalvir, 28850 Torrej\'on de Ardoz, Madrid, Spain\\
$^{4}$Universidad Autonoma de Madrid, Ciudad Universitaria de Cantoblanco, E-28049, Madrid, Spain\\
$^{5}$CIAFF, Facultad de Ciencias, Universidad Autonoma de Madrid, 28049 Madrid, Spain\\
$^{6}$Department of Astronomy, The University of Texas at Austin, 2515 Speedway, Stop C1400, Austin, TX 78712-1205, USA\\
$^{7}$International Associate, Cosmic Dawn Center (DAWN) \\
$^{8}$Instituto de Astrof\'\i sica de Canarias, C/V\'\i a L\'actea, s/n, E-38205 San Crist\'obal de La Laguna, Tenerife, Spain\\
$^{9}$Universidad de La Laguna, Dpto. Astrof\'\i sica, E-38206 San Crist\'obal de La Laguna, Tenerife, Spain\\
$^{10}$INAF – OAS, Osservatorio di Astrofisica e Scienza dello Spazio di Bologna, via Gobetti 93/3, I-40129 Bologna, Italy\\
}

\date{}

\begin{document}
\label{firstpage}
\pagerange{\pageref{firstpage}--\pageref{lastpage}}
\maketitle

\begin{abstract}

We present the discovery and analysis of J1316$+$2614 at $z=3.6130$, a UV-bright star-forming galaxy ($M_{\rm UV} \simeq -24.7$) with large escape of Lyman continuum (LyC) radiation. J1316$+$2614 is a young ($\simeq 10$~Myr) star-forming galaxy with $SFR \simeq 500$~$M_{\odot}$~yr$^{-1}$ and a starburst mass of log($M_{\star}/M_{\odot}) \simeq 9.7$. It shows a very steep UV continuum, $\beta_{\rm UV} \simeq -2.59 \pm 0.05$, consistent with residual dust obscuration, $E(B-V)\simeq 0$. LyC emission is detected with high significance ($\simeq 17 \sigma$) down to $830$\AA, for which a very high relative (absolute) LyC escape fraction $f_{\rm esc} \rm (LyC) \simeq 0.92$ ($\simeq 0.87$) is inferred. The contribution of a foreground or AGN contamination to the LyC signal is discussed, but is unlikely. J1316$+$2614 is the most powerful ionizing source known among the star-forming galaxy population, both in terms of production ($Q_{\rm H} \approx 10^{56}$~s$^{-1}$) and escape of ionizing photons ($f_{\rm esc} \rm (LyC) \approx 0.9$). 
Nebular emission in Ly$\alpha$, H$\beta$, and other rest-frame optical lines are detected, but these are weak ($EW_{0} \rm [H\beta] \simeq 35$\AA), with their strengths reduced roughly by $\simeq 90\%$. J1316$+$2614 is the first case known where the effect of large escape of ionizing photons on the strength of nebular lines and continuum emission is clearly observed. Gas inflows are detected in J1316$+$2614 from the blue-dominated peak Ly$\alpha$ emission (with a blue-to-red peak line ratio $I_{\rm blue}/I_{\rm red} \simeq 3.7$) and redshifted ISM absorption ($\simeq 100$~km\,s$^{-1}$). Our results suggest that J1316$+$2614 is undergoing a gas compaction event, possibly representing a short-lived phase in the evolution of massive and compact galaxies, where strong gas inflows have triggered an extreme star formation episode and nearly 100\% LyC photons are escaping.

\end{abstract}

\begin{keywords}
galaxies: formation -- galaxies: evolution -- galaxies: high-redshift 
\end{keywords}



\section{Introduction}

One of the key questions of modern observational cosmology is to identify the sources responsible for the ionization of the neutral intergalactic medium (IGM) in the first Gyr of cosmic time – the Epoch of Reionization (EoR). It is widely recognized from various observations that the EoR happened at $z \sim 6-15$ \citep[e.g.,][]{banados2018, mason2018, planck2020}, yet the sources that were responsible for the majority of ionizing photons (with $>$13.6\,eV; hereafter Lyman continuum, LyC) remain elusive.

Faint star-forming galaxies are thought to be the main drivers for reionization \citep[see: e.g.,][]{robertson2015, finkelstein2019} due to their high number density. Also, these sources may have higher ionizing photon production efficiency ($\xi_{\rm ion}$, e.g., \citealt{schaerer2016, maseda2020}) compared to UV-bright, massive counterparts. However, LyC surveys of faint star-forming galaxies at $z \lesssim 4$, where LyC radiation can be directly observed and measured, have revealed relatively low or negligible LyC escape fractions on average ($f_{\rm esc} \rm (LyC) \lesssim 0.1$; e.g., \citealt{marchi2017, rutkowski2017, fletcher2019, bian2020, flury2022a}). Only for a few individual galaxies LyC is detected with high significance and large LyC escape fractions are inferred ($f_{\rm esc} \rm (LyC) \geq 0.2$; e.g., \citealt{vanzella2016, debarros2016, izotov2018a, izotov2018b, steidel2018, fletcher2019, flury2022a}).

Dedicated surveys have been also carried out to investigate the properties of these sources and understand the main mechanisms for LyC leakage \citep[e.g.,][]{steidel2018, fletcher2019, flury2022a}. Despite these huge observational efforts, the connection between LyC leakage and different galaxy properties is still not fully understood. Some properties appear to correlate with LyC leakage. UV compact morphologies and large star-formation surface densities ($\Sigma \rm SFR$), low dust attenuation, low covering fraction of gas and high ionization parameters among others, appear to be common in strong LyC emitters \citep[e.g.,][]{jaskot2013, alexandroff2015, sharma2016, izotov2018a, flury2022b, saldana2022}. On the other hand, other properties appear to not correlate at all or show weak correlation only, including the stellar mass, metallicity or spectral hardness \citep[e.g.,][]{flury2022b, marques2022, saxena2022}. Motivated by these findings and other empirical trends, indirect tracers of LyC leakage have been proposed, tested, and some of them successfully established (e.g., \citealt{zackrisson2013, nakajima2014, verhamme2015, sharma2017, chisholm2018, izotov2018b, chisholm2022, saldana2022, schaerer2022, xu2022}). These can be used at all redshifts including in the EoR, where the detection of LyC is statistically unlikely due to the opacity of the IGM \citep{inoue2014}.

On the opposite side of the UV luminosity function are the more rare, UV-bright star-forming galaxies. By definition, these sources probe intense star-formation and, therefore, high production of ionizing photons. Recent works have found remarkably luminous star-forming galaxies at $z>6$ \citep[e.g.,][]{hashimoto2019, morishita2020, endsley2021, bouwens2022}, including the highest spectroscopically and photometrically redshift sources known, leading to important implications \citep[][]{oesch2016, harikane2022}. The volume density inferred for these luminous sources is much higher than that predicted by models \citep[e.g.,][]{mason2018}, by factors of $\sim 10-100$.  AGN contamination could be a natural explanation for the excess of UV-bright sources in the early Universe, but recent results disfavour such scenario \citep{finkelstein2022}. Rather than that, higher star-formation efficiency than previously thought at early times and/or lower dust obscuration towards the UV-bright end of luminosity functions are invoked \citep[e.g.,][]{yung2019, harikane2022b}.
Also, it has been shown that there is little or no evolution in the number density of very bright sources ($M_{\rm UV} \simeq -23$) between $z=4$ and $z=10$ \citep[e.g.,][]{bowler2020, harikane2022b}, implying that the relative number of UV-bright to UV-faint sources increases towards higher redshifts. However, the number of UV-bright sources is still scarce and their space density highly uncertain, ranging from $\lesssim 10^{-7}$ Mpc$^{-3}$ to some $10^{-6}$ Mpc$^{-3}$ for $M_{\rm UV} \simeq -23$ at $z\simeq 7-8$ \citep[e.g.,][]{calvi2016, bowler2020, rojasruiz2020, leethochawalit2022}, possibly reflecting cosmic variance effects. 
While a precise determination of the number density of these EoR bright sources can be done with upcoming very wide surveys like \textit{Euclid} and \textit{Nancy Grace Roman Space Telescope}, other important questions remain to be answered: what role do these UV-bright sources have to galaxy formation and evolution, and to cosmic reionization?

Recently, we have undertaken a search for very luminous star-forming galaxies at $z\gtrsim 2$ within the $\sim 9000$~deg$^{2}$-wide extended Baryon Oscillation Spectroscopic Survey \citep[eBOSS][]{abolfathi2018} of the Sloan Digital Sky Survey \citep[SDSS:][]{eisenstein2011}. The first results of this project were presented in \cite{marques2020b}, \cite{alvarez2021}, and \cite{marques2021}, where two UV-bright ($M_{\rm UV} < -24$) star-forming galaxies were analyzed in detail, BOSS-EUVLG1 at $z=2.47$ and J0121$+$0025 at $z=3.24$. These sources are very young ($\lesssim 10$~Myr) and compact ($r_{\rm eff} \sim 1$~kpc) starbursts with star-formation rates $\rm SFR \simeq 1000~M_{\odot}$~yr$^{-1}$, but with low dust attenuation ($E(B-V) \lesssim 0.1$). For example, dust is not detected in BOSS-EVULG1, yielding a dust mass log($M_{\rm dust} / M_{\odot}) < 7.3$ and a dust to stellar mass ratio $M_{\rm dust}/M_{\star} \simeq 2\times10^{-3}$ \citep{marques2020b}. 
While no direct information on the LyC leakage is known for BOSS-EUVLG1, J0121$+$0025 shows significant emission below the Lyman edge, compatible with large $f_{\rm esc} (LyC) \approx 40\%$ \citep{marques2021}, implying that that UV-bright galaxies can emit ionizing photons. The spectral energy distribution (SED) of these sources are fully dominated by the young starburst with specific SFR of $sSFR \sim 100$~Gyr$^{-1}$, and without a relevant old stellar component, possibly indicating that the bulk of their stellar mass ($\simeq 10^{10} M_{\odot}$) was assembled in a few Myr. All together, these authors speculate that these rare UV-bright starbursts ($\sim 10^{-9} \rm Mpc^{-3}$, \citealt{marques2020b}) could represent an early and short-lived phase in the evolution of massive galaxies, such as compact ellipticals or dusty star-forming galaxies found at $\simeq 2-3$. 
However, the main mechanism and physical conditions behind the formation of such extreme starbursts are not understood yet.

In this work, we present SDSS J131629.61$+$261407.0 at $z=3.6130$ ($\alpha$,~$\delta$ [J2000] = 199.1234$^{\circ}$, 26.2353$^{\circ}$, hereafter J1316$+$2614). J1316$+$2614 is a very luminous ($M_{\rm UV} = -24.6$) star-forming galaxy with $\simeq 90$\% LyC leakage and signatures of inflowing gas, shedding further light on the physical mechanism of galaxy formation of these peculiar UV-bright sources. The paper is structured as follows. The discovery and follow-up observations are presented in Section \ref{section2}. The analysis of the rest-frame UV spectroscopic observations, including the ionizing and non-ionizing spectra of J1316$+$2614, is presented in Section \ref{section3}. In Section \ref{section4} we discuss the properties of J1316$+$2614 and compare them with those from other sources in the literature. Finally, we present the summary of our main findings in Section \ref{conclusion}.
Throughout this work, we assume concordance cosmology with $\Omega_{\rm m} = 0.274$, $\Omega_{\Lambda} = 0.726$, and $H_{0} = 70$ km s$^{-1}$ Mpc$^{-1}$. Magnitudes are given in the AB system. Absolute magnitudes and luminosities are not corrected by dust.

\section{Discovery and Follow-up Observations}\label{section2}

J1316$+$2614 at $z=3.613$ was discovered as part of our search for luminous star-forming galaxies at high redshift ($z>2$) within the $\sim 9300$~deg$^{2}$-wide eBOSS/SDSS \citep{eisenstein2011, abolfathi2018}. Similar as BOSS-EUVLG1 and J0121$+$0025 \citep{marques2020b, marques2021}, J1316$+$2614 is classified as a QSO in the Data Release 14 Quasar catalog \citep[][]{paris2018}, but its BOSS spectrum (plate-mjd-fiberid: 5997-56309-375)\footnote{\url{http://skyserver.sdss.org/dr14/en/tools/explore/summary.aspx?id=1237667442439750118}} shows features characteristic of a young star-forming galaxy, without any hint of AGN activity. It shows narrow Ly$\alpha$ emission (a full width half maximum, FWHM, of $\simeq 500$~km~s$^{-1}$) and P-Cygni features in the wind lines  N~{\sc v} 1240\AA{ }and C~{\sc iv} 1550\AA, that is indicative of a young stellar population. Moreover, J1316$+$2614 shows a compact morphology without any evidence of being magnified by gravitational lensing, such as multiple images or arc-like morphologies, or the presence of nearby bright lens (Figure \ref{fig_image}).  

\subsection{Optical observations}

We obtained optical spectra of J1316$+$2614 with the Optical System for Imaging and low-Intermediate-Resolution Integrated Spectroscopy instrument (OSIRIS)\footnote{\url{http://www.gtc.iac.es/instruments/osiris/}} on the 10.4m Gran Telescopio Canarias (GTC) telescope as part of the programs GTCMULTIPLE2F-18A and GTC29-21A (PI: R.~Marques-Chaves). OSIRIS observations were performed under $\simeq0.9^{\prime \prime}-1.1^{\prime \prime}$ seeing conditions (FWHM) using the R2500R and R1000B grisms providing spectral resolutions $\rm R\sim 1800$ and $\rm R\sim 700$ and coverage of 5580-7700\AA{ }and 3600-7600\AA, respectively. Long-slits with 1.0$^{\prime \prime}$-width were centered on J1316$+$2614 and oriented with the parallactic angle (Figure \ref{fig_image}). Total on-source exposure times are 150~min and 60~min for the medium and low-resolution observations, respectively. Table \ref{tab0} summarizes the GTC spectroscopic observations. 

\begin{table}
\begin{center}
\caption{Summary of the GTC spectroscopic observations of J1316$+$2614. \label{tab0}}
\resizebox{0.47\textwidth}{!}{%
\begin{tabular}{l c c c c}
\hline \hline
\smallskip
\smallskip
Instrument & Grism ($R$) & Spec. range &  Exp. time & Date \\
 &  & ($\mu$m) &    (sec) &  \\
\hline 
OSIRIS & R1000B (700) & $0.360-0.750$  & $900 \times 4$  & 10 April 2018  \\
OSIRIS & R2500R (1800) & $0.558-0.769$ & $750 \times 12$  & 20 April 2021 \\
EMIR & HK  (700) & $1.454-2.405$  & $160 \times 16$  & 8 May 2018 \\
\hline 
\end{tabular}}
\end{center}
\end{table}

Data were reduced following standard reduction procedures using {\sc Iraf}. These include subtraction of the bias and further correction of the flat-field. The wavelength calibration is done using HgAr+Ne+Xe arc lamps data. 2D spectra are background subtracted using sky regions around J1316$+$2614. Individual 1D spectra are extracted, stacked and corrected for the instrumental response using observations of the standard star Ross 640. We use the extinction curve of \cite{cardelli1989} and the extinction map of \cite{schlafly2011} to correct for the reddening effect in the Galaxy. Finally, the flux of the spectrum is matched to that obtained from photometry in the $R$-band to account for slit-losses, and corrected for telluric absorption using the {\sc Iraf} \textit{telluric} routine. 

\subsection{Near-IR observations}

\begin{figure}
  \centering
  \includegraphics[width=0.48\textwidth]{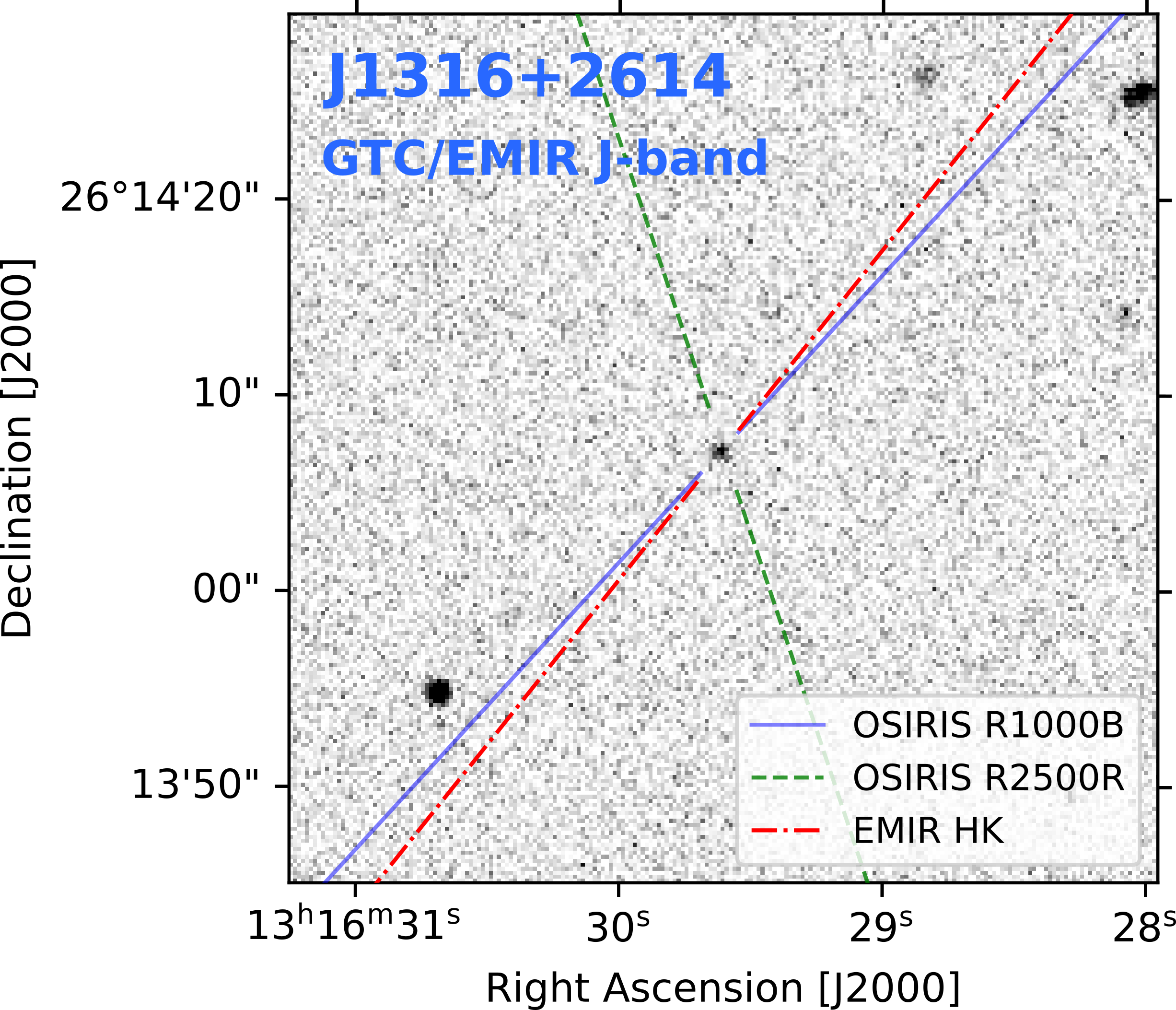}
  \caption{Cutout of J1316$+$2614 from the GTC/EMIR $J$-band image. J1316$+$2614 is located in the center of the image. The orientation of the GTC long-slits are marked in blue (solid), green (dashed) and red (dash-dot) for the optical OSIRIS low-resolution (R1000B), medium-resolution (R2500R) and near-IR EMIR (HK) grisms, respectively.}
  \label{fig_image}
\end{figure}

Near-IR spectra were obtained with the Espectr\'ografo Multiobjeto Infra-Rojo (EMIR)\footnote{\url{http://www.gtc.iac.es/instruments/emir/}} on the GTC under good seeing conditions ($0.7^{\prime \prime}$ FWHM). The $HK$ grism with a $0.8^{\prime \prime}$-width was used with a total observing time of 43~min with a standard 10$^{\prime \prime}$ ABBA dither, providing a spectral resolution $\rm R\sim 700$ and coverage of 1.45-2.41$\mu$m. Reduction of the near-IR spectrum was performed using the official EMIR pipeline\footnote{\url{https://pyemir.readthedocs.io/en/latest/index.html}}. Spectra were flux calibrated using a standard star observed in that night, and fluxes matched to those obtained from photometry. 
We also obtained near-IR imaging with EMIR using the $Y$, $J$, $H$, and $K_{\rm s}$ filters (Figure~\ref{fig_image}). Total exposure times range from $\simeq 10$~min to $\simeq 30$min. Images were also reduced using the EMIR pipeline and were flux calibrated against 2MASS stars in the field.
J1316$+$2614 is detected with high significance in all near-IR bands and appears unresolved in these images with seeing conditions $\simeq 0.9^{\prime \prime}-1.1^{\prime \prime}$ FWHM.

\section{Results}\label{section3}

\subsection{Rest-frame UV spectrum and young stellar population}\label{section31}

\begin{figure*}
  \centering
  \includegraphics[width=0.99\textwidth]{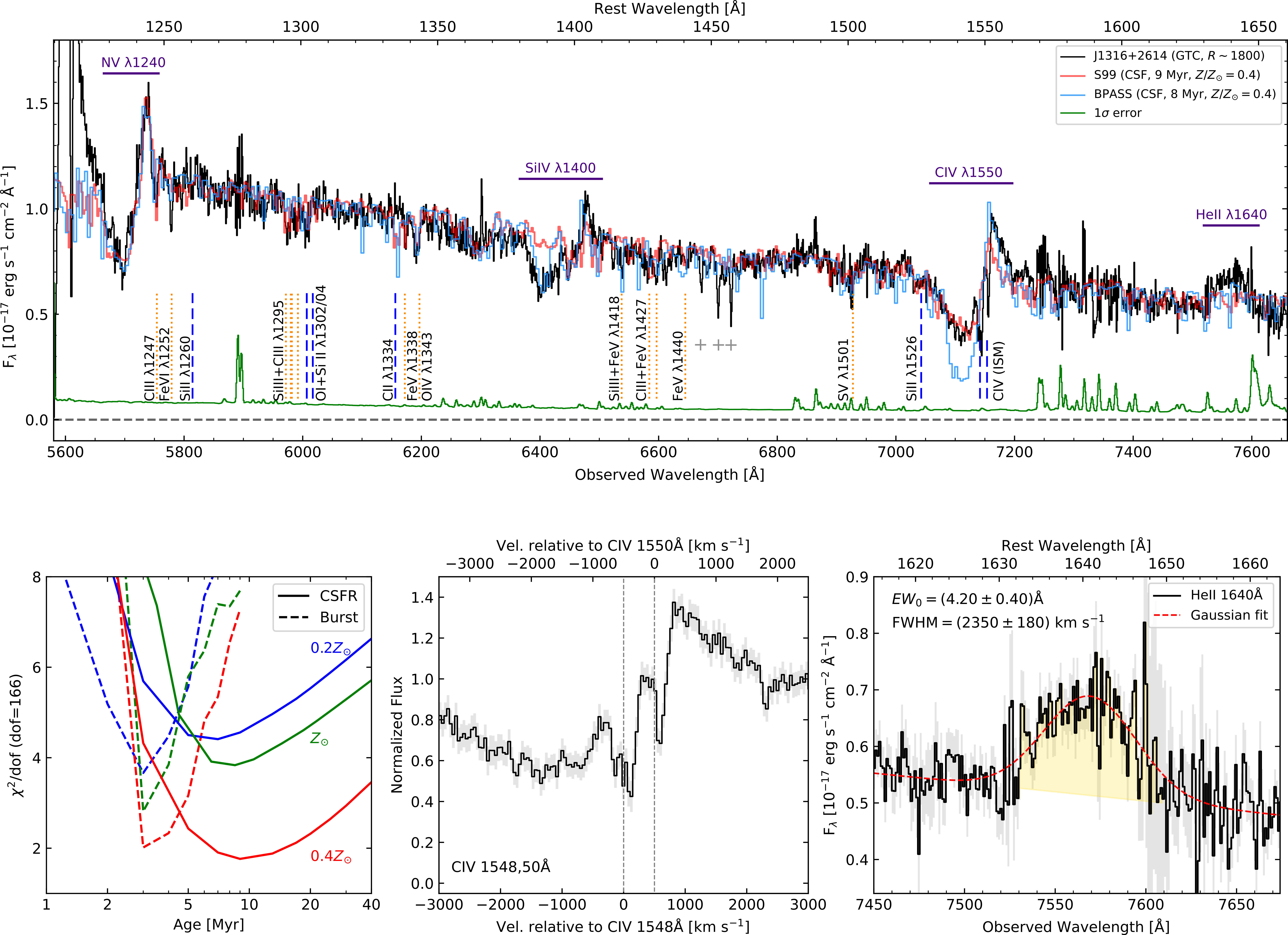}
  \caption{Top panel: GTC/OSIRIS rest-frame UV spectrum of J1316$+$2614 (black) and its corresponding $1\sigma$ uncertainty (green). Vertical lines marked below the spectrum identify the position of low-ionization ISM absorption (blue) and several photospheric absorption lines (yellow), some of them resolved and detected with high significance for which the systemic redshift was derived, $z_{\rm sys} = 3.6130 \pm 0.0008$. The spectrum also shows three absorption lines that not related to J1316$+$2614 (cross symbols), two of them originated from an intervening system at $z\simeq 2.579$ (Al~{\sc iii}~1854,1863\AA). The best-fit Starburst99 model with a continuous SFR over an age of 9~Myr and $Z_{\star}/Z_{\odot} =0.4$ is shown in red. For comparison, a BPASS model with similar stellar properties as the best-fit Starburst99 model is also shown (blue). 
  Bottom left shows the results of our $\chi^{2}$ minimization using continuous SFR (solid) and burst (dashed) Starburst99 models and different metallicities. Bottom middle shows a zoom-in into the C~{\sc iv} profile. Two narrow absorption lines associated with the ISM components of C~{\sc iv} (at $\simeq 1548$\AA{ }and $\simeq 1550$\AA) are detected and are redshifted by $\simeq 100$~km~s$^{-1}$ relative to the systemic velocity.
  The bottom right panel shows the He~{\sc ii} emission seen in J1316$+$2614 (black) and the Gaussian fit (red), with the fitted parameters, $EW_{0}$ and FWHM, indicated in the legend.}
  \label{fig1}
\end{figure*}

The OSIRIS/GTC medium-resolution spectrum of J1316$+$2614 is shown in Figure \ref{fig1}. It shows a blue UV continuum with a slope, $-2.59 \pm 0.05$, measured using two spectral windows of 30\AA-width at 1300\AA{ }and 1600\AA{ }rest-frame and assuming that $F_{\lambda} \propto \lambda^{\beta_{\rm UV}}$. This is steeper, but consistent within the uncertainties, than the slope derived from the photometry using $i$- and $J$-bands, $\beta_{\rm UV}^{\rm phot} = -2.43 \pm 0.17$, that probes slightly longer wavelengths of $\simeq 1600$\AA{ }to $\simeq 2700$\AA{ }(rest-frame), respectively. 

The most prominent features in the UV spectrum of J1316$+$2614 are the strong P-Cygni profiles seen in stellar wind lines. In particular, the profiles of N~{\sc v}~1240\AA{ }and C~{\sc iv}~1550\AA{ }are known to be strongly dependent on the age and metallicity of the stellar population \citep[e.g.,][]{chisholm2019}. To investigate these properties, we compare the observed N~{\sc v} and C~{\sc iv} profiles to those obtained with the spectral synthesis code {\sc Starburst99} \citep[S99:][]{leitherer1999}, following the same methodology described in \cite{marques2018} and \cite{marques2020}. 
S99 models are preferred in this fit instead of BPASS ones \citep{stanway2016}, because the shape of the P-Cygni absorption of C~{\sc iv} in the later is difficult to match with observations (see discussion in \citealt{steidel2016}). The effect of binary stars contributing to the spectrum of J1316+2614 will be discussed in Section~\ref{section43}. 

We generate UV spectra using standard Geneva tracks with a grid of metallicities ($Z_{\star}/Z_{\odot}$, where we assume $Z_{\odot}=0.02$) of 0.05, 0.2, 0.4 and 1. We assume both continuous and burst star-formation histories and a initial mass function with a power slope index $\alpha=-2.35$ over the mass range $0.5 < M_{\star}/M_{\odot} < 100$. 
A $\chi^{2}$ minimization is performed in the continuum normalized profiles of N~{\sc v} and C~{\sc iv} over the spectral ranges $1227-1246$\AA{ }and $1526-1545$\AA, respectively. For C~{\sc iv}, we avoid fitting regions $>1545$\AA{ }as they include the interstellar component of C~{\sc iv} that is redshifted by $\simeq 100$~km~s$^{-1}$ (see Figure~\ref{fig1} and Section~\ref{section311}) and possibly nebular emission. 

P-Cygni profiles of N~{\sc v} and C~{\sc iv} are best reproduced by a S99 model with a continuous star-formation history with an age of $9$~Myr and $Z_{\star}/Z_{\odot} \simeq 0.4$ with a $\chi^{2}/N_{\rm dof} = 1.76$ (red in the left bottom panel of Figure \ref{fig1}), where the number of degree of freedom is $N_{\rm dof} = 166$. S99 models with $Z_{\star}/Z_{\odot} = 0.2$ (and $Z_{\star}/Z_{\odot} = 1.0$) over-predict (under-predict) the strength of C~{\sc iv} profile, increasing the $\chi^{2}$ by a factor $\gtrsim 2.3$. Similarly, S99 models with ages $<4$~Myr (and $>20$~Myr) will over-predicted (under-predicted) the strength of N~{\sc v} profile. On the other hand, if a burst star-formation history is assumed, a fairly good fit is found for a burst age of $3-4$~Myr, but the fit is still worse ($\chi^{2}/N_{\rm dof} = 2.0$) than the one assuming a continuous star-formation history. Therefore, we consider from now on a continuous star-formation history with an age of $9\pm 5$~Myr and $Z_{\star}/Z_{\odot} \simeq 0.4$ as the best fit of J1316$+$2614. The ionizing production efficiency of this model, defined as the number of ionizing photons produced per unit UV luminosity ($\xi_{\rm ion} = Q_{\rm H} / L_{\rm UV, int}$), is log($\xi_{\rm ion} \rm [erg^{-1} Hz]) = 25.40$. This is higher than the $\xi_{\rm ion}$ inferred for typical high-$z$ LBGs (log($\xi_{\rm ion}/\rm erg^{-1}Hz) \simeq  24.8$, e.g., \citealt{bouwens2016b}), but comparable with that inferred for LAEs (e.g., \citealt{matthee2017b}).

We also take in consideration the dust attenuation. The observed $\beta_{\rm UV} = -2.59 \pm 0.05$ in J1316$+$2614 is similar than the intrinsic UV slope inferred for the best-fit S99 model, $\beta_{\rm UV} \rm (S99) = -2.62$, i.e., without considering the effect of dust or the inclusion of nebular continuum. Therefore, the observed slope is consistent with almost zero attenuation, $E (B-V)_{\star} = 0.006^{+0.018}_{-0.006}$, assuming the \citealt{calzetti2000} extinction curve. 

In addition to the N~{\sc v} and C~{\sc iv} P-Cygni profiles, other stellar features are detected and their strengths are relatively well reproduced by the best-fit S99 model. 
The photospheric absorption in C~{\sc iii}~1247\AA, S~{\sc v}~1501\AA, and some blends from multiple transitions such as S~{\sc iii} and C~{\sc iii} around $\sim$1295\AA{ }and by Fe~{\sc v} and O~{\sc iv} around $\sim$1340\AA{ }are well detected (see Figure \ref{fig1}) and their strengths are consistent with the best-fit S99 model. As these lines arise from stellar atmospheres, we use them to determine the systemic redshift of J1316$+$2614 as $z_{\rm sys} = 3.6130 \pm 0.0008$.

Lastly, we note the poor agreement of the best fit S99 model in two important features, Si~{\sc iv}~1393,1402\AA{ }and He~{\sc ii}~1640\AA. A strong P-Cygni contribution is detected in Si~{\sc iv}~1393,1402\AA, but the best-fit S99 model does not reproduce its strength.  
Such profiles are usually seen in some early-O type supergiants,  Wolf-Rayet (W-R) stars (e.g., \citealt{garcia2004, crowther2016}) or in metal-rich stars, and might indicate different (lower) effective temperatures. Only S99 models with $Z/Z_{\odot} \geq 1$ can show Si~{\sc iv} profiles with similar strength as the one observed in J1316$+$2614. However, these high metallicity models are disfavored as they predict much stronger absorption in photospheric lines than the observed ones. It is worth to note that BPASS models with similar metallicity as our best-fit S99 model (i.e., $0.4Z_{\odot}$) can predict relatively well the observed P-Cygni profile in Si~{\sc iv} (shown in blue in Figure~\ref{fig1} for comparison). 
The best-fit S99 model also fails to reproduce the strength of He~{\sc ii}~1640\AA{ }emission in J1316$+$2614 (Figure \ref{fig1}). Fitting a Gaussian profile to the He~{\sc ii} line, we measure a line width $\rm FWHM = 2350 \pm 180$~km~s$^{-1}$ and a rest-frame equivalent width $EW_{0}=4.2 \pm 0.4$\AA. This is much stronger than that found in typical star-forming galaxies ($EW_{0}^{\rm LBGs} \simeq 1.3$\AA, e.g., \citealt{shapley2003}), and may indicate a significant contribution of stars with mass $>100 M_{\odot}$ (\citealt{senchyna2021, martins2022}) that show very strong and broad He~{\sc ii} emission \citep[e.g.,][]{crowther2016}. 
A more detailed investigation on the origin of the strong He~{\sc ii} in J1316$+$2614 and in other luminous star-forming galaxies will be presented in a forthcoming work (Upadhyaya et al. in prep.).

\subsection{ISM absorption lines}\label{section311}

Interstellar medium (ISM) absorption lines appear very weak in the spectrum of J1316$+$2614. We checked the presence of the main low-ionization ISM (LIS) lines that are typically observed in many other star-forming galaxies (e.g., \citealt{shapley2003}). These include Si~{\sc ii}~1260\AA, O~{\sc i}+Si{\sc ii}~1302\AA, C~{\sc ii}~1334\AA, or Si~{\sc ii}~1526\AA{ }(blue dashed lines in Figure~\ref{fig1}). However, these lines are not detected in the spectrum of J1316$+$2614 despite the high SNR of the continuum ($\simeq 15-20$~pix$^{-1}$). We only detect a faint absorption in O~{\sc i}+Si~{\sc ii} at the systemic velocity with $EW_{0} \simeq 1$\AA, but stellar models can explain its strength (Figure~\ref{fig1}).

On the other hand, two narrow absorption lines are detected at $\simeq 7142$\AA{ }and $\simeq 7154$\AA{ }and are consistent to be the ISM components of the C~{\sc iv} doublet (1548\AA{ }and 1550\AA, see bottom panel of Figure \ref{fig1}). 
The lines present narrow profiles being almost unresolved (FWHM $\lesssim 200$~km~s$^{-1}$). They show $EW_{0} \simeq (1 - 3)$\AA, although the uncertainties are large given the difficulty to estimate the continuum level within the P-Cygni profile. Interestingly, the C~{\sc iv} ISM absorption lines are redshifted by $\simeq 100$~km~s$^{-1}$ relative to the systemic velocity derived using stellar absorption lines, indicating that the high-ionized ISM is apparently moving towards the young stars.

\begin{figure*}
  \centering
  \includegraphics[width=0.99\textwidth]{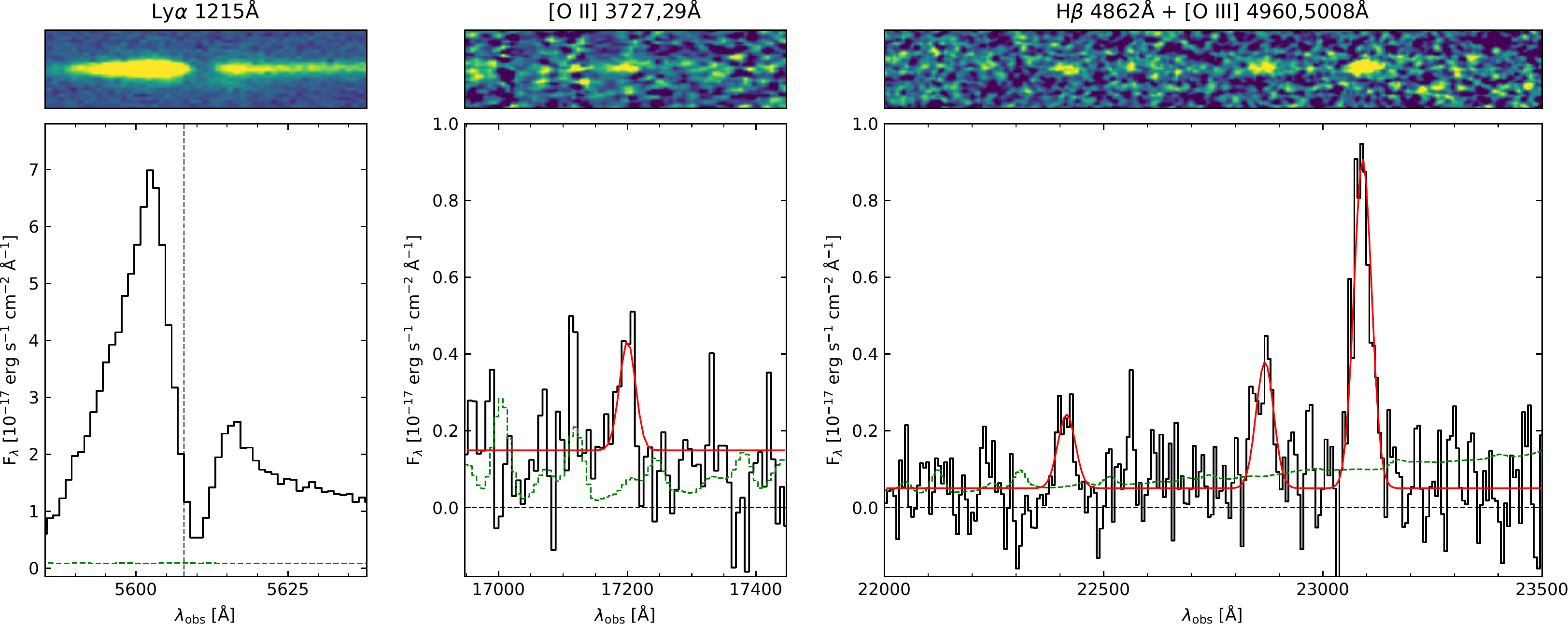}
  \caption{Nebular emission detected in J1316$+$2614. Left panel shows the Ly$\alpha$ emission (black and 1$\sigma$ error in green) seen in the medium resolution OSIRIS spectrum. It shows a double peak emission around the systemic velocity (vertical dashed line). The blue peak is more intense than the red one, suggestive of neutral gas infall. Middle and right panels show the near-IR spectrum with EMIR around the regions of [O~{\sc ii}] and H$\beta$+[O~{\sc iii}], respectively. 1$\sigma$ error is shown in green and the performed Gaussian fits for each line are shown in red.  }
  \label{fig2}
\end{figure*}

\subsection{Nebular emission}\label{section32}

J1316$+$2614 shows significant ($>5\sigma$) emission in several nebular lines, such as Ly$\alpha$~1215\AA, H$\beta$~4862\AA, and in the [O~{\sc iii}] doublet at 4960\AA{ }and 5008\AA. These are shown in Figure \ref{fig2}. 

The Ly$\alpha$ line shows two peaks around the systemic velocity, with the blue peak more intense than the red one. We measure a ratio between the blue and red emission $I_{\rm B}/I_{\rm R} = 3.7 \pm 0.1$ and a velocity peak separation $\Delta v \rm (Ly\alpha) = 680 \pm 70$~km~s$^{-1}$. Fitting a Gaussian profile to the blue and red emission, we find line widths of $440$~km~s$^{-1}$ and $510$~km~s$^{-1}$ (FWHM, and corrected for instrumental broadening), respectively. We note, however, that these lines do not show a symmetric profile and, therefore, are not well reproduced by Gaussian profiles. Because the continuum level around Ly$\alpha$ is not easy to constrain (due to the IGM absorption and the N~{\sc v}~1240\AA{ }P-Cygni profile at shorter and longer wavelengths, respectively), we subtract and divide our spectrum by the best fit S99 model. We then measure respectively the flux and equivalent width of the nebular component of Ly$\alpha$ by integrating its flux within the spectral region $\pm 1000$~km~s$^{-1}$ around the systemic velocity. 
Ly$\alpha$ has a total flux of $F \rm (Ly\alpha) = (9.62 \pm 0.02) \times 10^{-16}$~erg~s$^{-1}$~cm$^{-2}$, and $EW_{0} = 20.5\pm1.9$\AA, corresponding to a luminosity log$_{10}$($L_{\rm Ly \alpha} / \rm erg~s^{-1}) = 44.09\pm0.10$. These values are consistent with those measured using the BOSS spectrum, which uses a larger aperture ($2^{\prime \prime}$-diameter fiber) than OSIRIS ($1^{\prime \prime}$-width), suggesting that there is little scatter at larger radii. 

It is worth to note that Ly$\alpha$ profiles like this, i.e., showing the blue peak more intense than the red one, are extremely rare. It has been observed only in a few star-forming galaxies \citep[][]{wofford2013, erb2014, trainor2015, izotov2020, furtak2022} and in other type of sources, such as AGNs or extended Ly$\alpha$ nebulae \citep[][]{martin2015, vanzella2017b, ao2020, daddi2021, li2022}. This profile is usually linked to radial infall of gas \citep{dijkstra2006}. 
Other scenarios, such as emission from two kinetically different sources (e.g., merger) or disc rotation, are unlikely. In such cases, similar doubled peaked profiles should be present in other nebular lines, which is not the case. As described next, rest-frame optical lines in J1316$+$2614 do not show this complex profile and can be  well modelled with single Gaussian profiles. 
We note however that rest-frame optical lines are observed with much lower spectral resolution and SNR than Ly$\alpha$, which makes it difficult a direct comparison of their profiles.

Turning to the rest-frame optical lines, three Gaussian profiles are used to fit simultaneously H$\beta$ and [O~{\sc iii}]~4960,5008\AA, and a constant (in $f_{\lambda}$) for the continuum. We measure $F \rm (H\beta) = (9.1\pm1.4) \times 10^{-17}$~erg~s$^{-1}$~cm$^{-2}$, $F \rm ([OIII]~4960) = (15.5 \pm 2.2) \times 10^{-17}$~erg~s$^{-1}$~cm$^{-2}$ and $F \rm ([OIII]~5008) = (42.5 \pm 3.8) \times 10^{-17}$~erg~s$^{-1}$~cm$^{-2}$. The lines appear barely resolved only, with $\rm FWHM \sim 380$~km~s$^{-1}$ after correcting for the EMIR instrumental broadening ($\simeq 450$~km~s$^{-1}$).
The continuum in the $K$-band is detected with a flux density corresponding to $m(K) = 21.42 \pm 0.13$ mag. We find rest-frame equivalent widths for H$\beta$ and [O~{\sc iii}]~4960\AA{ }and [O~{\sc iii}]~5008\AA{ }of $34.7\pm6.8$\AA, $59.3 \pm 10.8$\AA, and $162.0\pm23.9$\AA, respectively. 
Emission in nebular [O~{\sc ii}]~3727,3729\AA{ }is also detected around $\lambda \simeq 1.72\mu$m, but with low significance ($\simeq 2.5\sigma$) with a flux of $F \rm ([OII]) = (8.9\pm3.6) \times 10^{-17}$~erg~s$^{-1}$~cm$^{-2}$. Table \ref{tab2} summarizes our measurements.

Assuming case B recombination and $T=10^{4} K$, the intrinsic ratio between Ly$\alpha$ and H$\beta$ is 24.8. Using the observed fluxes of Ly$\alpha$ and H$\beta$ and considering $E(B-V)=0$, we infer a Ly$\alpha$ escape fraction $f_{\rm esc} \rm (Ly\alpha) = 0.43 \pm 0.12$.
We measure the line ratio $R23 = \rm ([OII + [OIII])/H \beta = 7.1 \pm 1.8$, for which we measure a nebular abundance 12+log(O/H)=$8.45\pm0.12$ ($\simeq 0.5 Z_{\odot}$) following \cite{pilyugin2005}. This is compatible with the stellar metallicity inferred for the young stellar population. A line ratio [O~{\sc iii}]/[O~{\sc ii}] of $O_{32}=4.8 \pm 2.1$ is also inferred and will be discussed in more detail in Section~\ref{section424}. Note that, as discussed in Section \ref{section42}, the high LyC escape fraction in J1316$+$2614 (Section \ref{section34}) is reducing the observed intensity of the nebular emission, and the suppression factor  might be different for recombination (Ly$\alpha$, H$\beta$) and forbidden metal lines ([O~{\sc ii}], [O~{\sc iii}]). This means that the nebular abundance inferred using $R23$ might not be valid and should be treated with caution. 

\begin{table}
\begin{center}
\caption{Nebular emission in J1316$+$2614. \label{tab2}}
\begin{tabular}{l c c}
\hline \hline
\smallskip
\smallskip
Line & $EW_{0}$ & Flux  \\
 & [\AA] & [$10^{-17}$~erg~s$^{-1}$~cm$^{-2}$] \\
\hline 
Ly$\alpha$~1215\AA & $20.5\pm1.9$ & $96.2 \pm 2$ \\

[O~II]~3727,3729\AA & $12.9\pm5.0$ & $ 8.9\pm3.6$ \\

H$\beta$~4862\AA & $34.7\pm6.8$ & $9.1\pm1.4$ \\

[O~III]~4960\AA & $59.3 \pm 10.8$ & $ 15.5 \pm 2.2$ \\

[O~III]~5008\AA& $162.0\pm23.9$ & $42.5 \pm 3.8$ \\

\hline 
\end{tabular}
\end{center}
\end{table}

\subsection{Lyman Continuum emission}\label{section33}

\begin{figure*}
  \centering
  \includegraphics[width=0.99\textwidth]{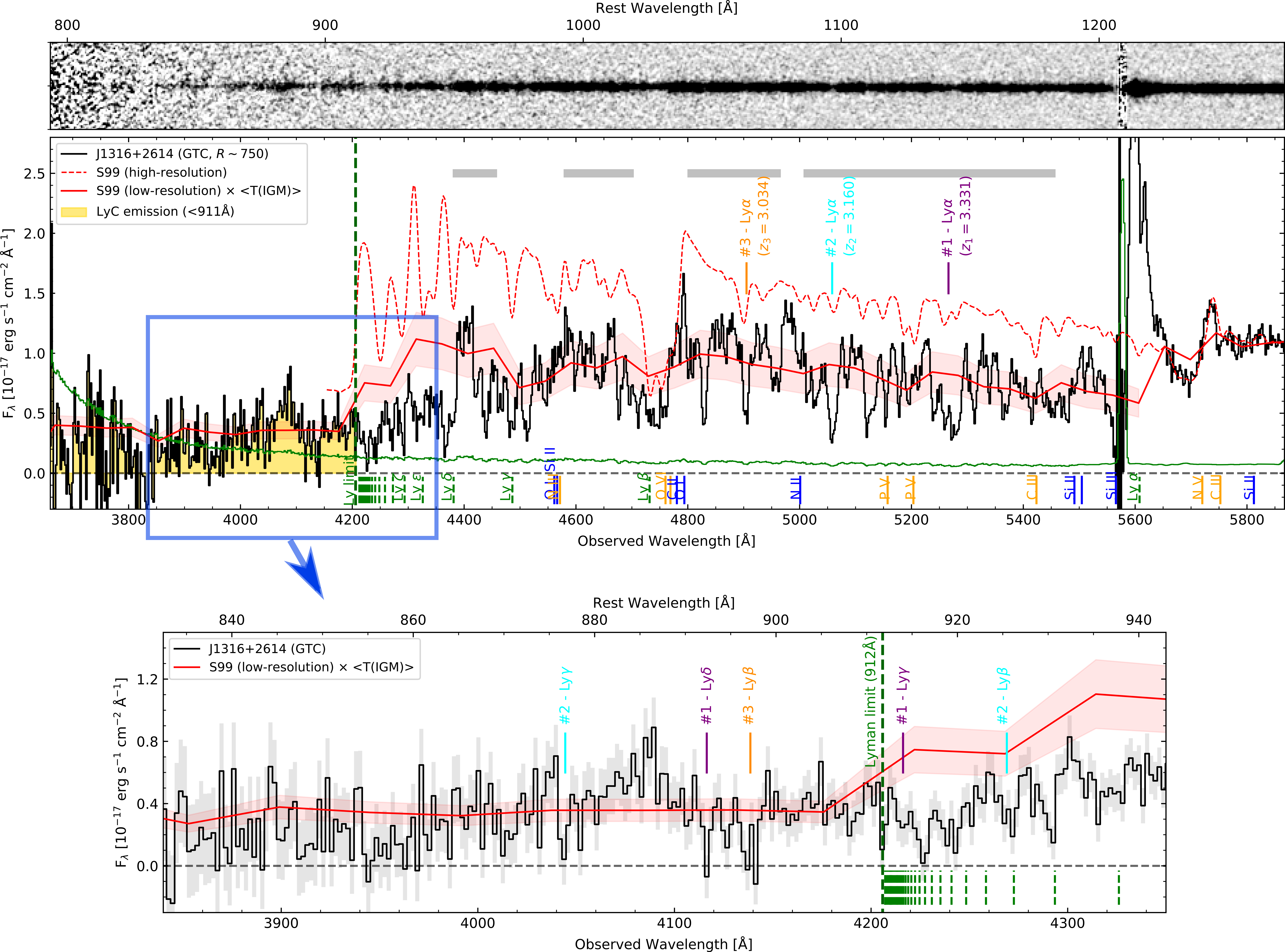}
  \caption{2D and 1D GTC low-resolution spectrum of J1316$+$2614 (black) and $1\sigma$ uncertainty (green). The spectral region in yellow corresponds to the emission below $\lambda_{0} < 912$\AA{ }, that is related to LyC emission. The position of the Lyman limit is marked with a green dashed line. The high and low-resolution best-fit S99 models (9~Mry age, $Z_{\star}/Z_{\odot}=0.4$ and $E(B-V)_{\star} = 0.0$) are shown in dashed and solid red, respectively. The later is corrected for the IGM transmission (<$T(IGM)$> $=0.59 \pm 0.17$). Horizontal grey lines mark the spectral windows used to infer $T(IGM)$, probe H~{\sc i} absorbers from the Lyman forest in the line of sight only and excluding other regions associated with the Lyman series and ISM from J1316$+$2614 (marked as green and blue vertical lines, respectively, below the 1d spectrum). Vertical yellow lines mark the position of stellar features associated with J1316$+$2614. Vertical lines above the spectrum mark the position of three strong H~{\sc i} absorbing systems identified at $z =  3.331$, $3.160$, and $3.034$ (\#1 to \# 3, respectively). The bottom panel shows a zoom-in to the LyC region.}
  \label{fig3}
\end{figure*}

The 40min low-resolution spectrum of J1316$+$2614 is shown in Figure \ref{fig3} and presents significant emission below $\lambda_{\rm obs} \simeq 4200$\AA, i.e. $\lambda_{\rm 0} < 911.8$\AA, that can be associated with LyC leakage. The emission is clearly detected down to rest-frame wavelengths $\simeq 830$\AA{ }(rest) with a total SNR $\simeq 17$ from $830 - 910$\AA{ }and an average SNR per spectral bin of $\simeq 1.6$~pix$^{-1}$.
We measure a flux density in the spectral region between 830\AA{ }and 910\AA{ }(rest) of $f_{870} (\rm obs) = 1.69\pm0.10 \mu$Jy, corresponding to a magnitude of $m_{870} = 23.33 \pm 0.06$~(AB). This yields a ratio of the ionizing to non-ionizing flux density $(f_{870} / f_{1500})_{\rm obs} = 0.146 \pm 0.011$ or $\Delta m = 2.1$~(AB).

The relative LyC photon escape fraction along the line of sight, $f_{\rm esc, rel}$~(LyC), is inferred by comparing the observed ratio of the ionizing to non-ionizing flux density of J1316$+$2614, ($f_{870}/f_{1500}$)$^{\rm obs}$, to that from the best-fit S99 model derived in Section \ref{section31}, ($f_{870}/f_{1500}$)$^{\rm S99}$, using the following formulation:

\begin{equation}\label{eq1}
    f_{\rm esc, rel} (LyC) = \frac{(f_{870}/ f_{1500})^{obs}} {(f_{870}/f_{1500})^{S99}} \times \frac{1}{T(IGM)},
\end{equation}

\noindent
where $T(IGM)$ is the IGM transmission, and flux ratios refer to $f_{\nu}$. Using the same rest-frame spectral windows of $830 - 910$\AA{ } and $1490-1510$\AA{ }, we infer $(f_{870} / f_{1500})^{\rm S99} = 0.27$. This means that $f_{\rm esc, rel} (LyC) = 0.54 / T(IGM)$, which implies that $T(IGM)$ should be at least larger than $>0.54$ to keep a physical $f_{\rm esc, rel} (LyC) \leq 1$ \citep[e.g.,][]{vanzella2012}. This also implies that $f_{\rm esc, rel}$~(LyC) must be $\geq 0.54$, where the most extreme value ($0.54$) stands for a completely transparent IGM ($T(IGM) =1.0$). 

A precise estimate of $T(IGM)$ below $<912$\AA\ (thus $f_{\rm esc} \rm (LyC)$) is not possible due to the stochastic nature and large fluctuation of the attenuation in one single line-of-sight \citep[e.g.,][]{inoue2008, inoue2014}. To overcome this, we infer $T(IGM)$ in the non-ionizing part of the spectrum of J1316$+$2614, from $912$\AA\ to $1215$\AA, and assume that is similar at $\lambda_{0} <912$\AA. 
Following \cite{marques2021}, we compare the flux of J1316$+$2614 to that of the best-fit S99 model in several spectral regions from $912 - 1215$\AA{ }(marked in Figure \ref{fig3} as grey horizontal bars), probing only H~{\sc i} absorbers from the Lyman forest. By doing this, we are excluding spectral regions associated with the Lyman series, ISM and CGM from J1316$+$2614, which are not included in S99 models. We derive a mean value and standard deviation $T(IGM) =0.59\pm0.17$, and assume this values for $\lambda_{0} <912$\AA. The inferred $T(IGM)$ in J1316$+$2614 is significantly larger than those obtained by other works using Monte Carlo simulations of the IGM transmission ($<T(IGM)> \simeq 0.2$ at $z\sim 3.5$, e.g., \citealt{steidel2018}). This suggests that the large flux density observed in J1316$+$2614, $f_{870} (\rm obs)$, is due to the combination of a large $f_{\rm esc, rel}$~(LyC) and a favourable IGM transmission. In Section \ref{section43} we discuss the implications of a favorable IGM transmission and other uncertainties regarding the adopted stellar model.

Using Equation \ref{eq1} we infer $f_{\rm esc, rel}$~$\rm (LyC) \sim 0.92^{+0.08}_{-0.20}$. We note that this value refers to the LyC escape fraction measured between $830 - 910$\AA{ }and along the line of sight. The uncertainties on $f_{\rm esc, rel}$~$\rm (LyC)$ refer to those arising from $T(IGM)$. Other sources of uncertainty should be less relevant or they are difficult to quantify. For example, the properties of the young stellar population, namely the age and metallicity, are relatively well constrained from the modeling of the stellar wind profiles (see Section \ref{section31}). Their uncertainties should not impact significantly on the inferred $f_{\rm esc, rel}$~$\rm (LyC)$ (see more discussion in Section \ref{section43}), but we also note they are model dependent. On the observational side, uncertainties due to the flux calibration or differential slit-losses between $\lambda_{0} < 912$\AA{ }and $\lambda_{0} \simeq 1500$\AA{ }are difficult to constrain. Nevertheless, the low-resolution GTC spectrum matches very well to that of BOSS spectrum, where LyC is also detected.

Assuming $E (B-V) = 0.006$, the absolute LyC escape fraction is $f_{\rm esc, abs} (LyC) = 0.87^{+0.08}_{-0.20}$. Given this, we estimate the number of ionizing photons escaping from J1316$+$2614, $Q^{\rm esc}_{H}$, by:

\begin{equation}\label{photon}
    Q^{\rm esc}_{H} =  f_{\rm esc, abs} (LyC) \times Q^{\rm int}_{H},
\end{equation}

\noindent
where $Q^{\rm int}_{H} = (8.5 \pm 1.7) \times 10^{55}$~s$^{-1}$ is the intrinsic ionizing photon production rate from the best-fit S99 model scaled to the absolute magnitude $M_{\rm UV} = -24.7$ of J1316$+$2614. We infer $Q^{\rm esc}_{H} = (7.3 \pm 2.0) \times 10^{55}$~s$^{-1}$.

\subsection{Multi-wavelength Spectral Energy Distribution}\label{section34}

In this Section we perform analysis to the multi-wavelength photometry and Spectral Energy Distribution (SED) of J1316$+$2614. Optical photometry in $g$, $r$, $i$, and $z$ bands are retrieved from SDSS using the {\sc model} photometry. 
For the near-IR images ($Y$, $J$, $H$, and $K_{\rm s}$) we use aperture photometry with a diameter of 2.5$\times$FWHM. J1316$+$2614 is not resolved in the optical (SDSS) nor near-IR (EMIR/GTC) with seeing conditions $\simeq 0.9^{\prime \prime}-1.2^{\prime \prime}$~FWHM. J1316$+$2614 is not detected in the mid-IR in the WISE2020 Catalog \citep[][]{marocco2021} at 3.4$\mu$m ($W1$) and 4.6$\mu$m ($W2$), with $W1 > 21.45$ and $W2 > 20.84$ at $5\sigma$. Table \ref{table1} summarizes the optical to mid-IR photometry of J1316$+$2614.

\begin{table}
\begin{center}
\caption{Optical to Mid-IR photometry of J1316$+$2614. \label{table1}}
\begin{tabular}{l c c c}
\hline \hline
\smallskip
\smallskip
Band & $\lambda_{\rm eff}$ & Magnitude & Telescope\\
 & ($\mu$m) & (AB)  & \\
\hline 
$g$ & 0.47 & $22.08\pm 0.08$ & SDSS \\
$r$ & 0.61 & $21.18\pm 0.05$ & SDSS \\
$i$ & 0.77 & $21.24\pm 0.08$ & SDSS \\
$z$ & 0.89 & $21.27\pm 0.31$ & SDSS \\
$Y$ & 1.01 & $21.38\pm 0.14$ & EMIR/GTC \\
$J$ & 1.25 & $21.48\pm 0.09$ & EMIR/GTC \\
$H$ & 1.61 & $21.70\pm 0.14$ & EMIR/GTC \\
$K_{\rm s}$ & 2.14 & $21.31\pm 0.07$ & EMIR/GTC \\
$W1$ & 3.35 & $>21.45$ (5$\sigma$) & WISE \\
$W2$ & 4.60 & $>20.84$ (5$\sigma$) & WISE \\
\hline 
\end{tabular}
\end{center}
\end{table}

\begin{figure*}
  \centering
  \includegraphics[width=0.99\textwidth]{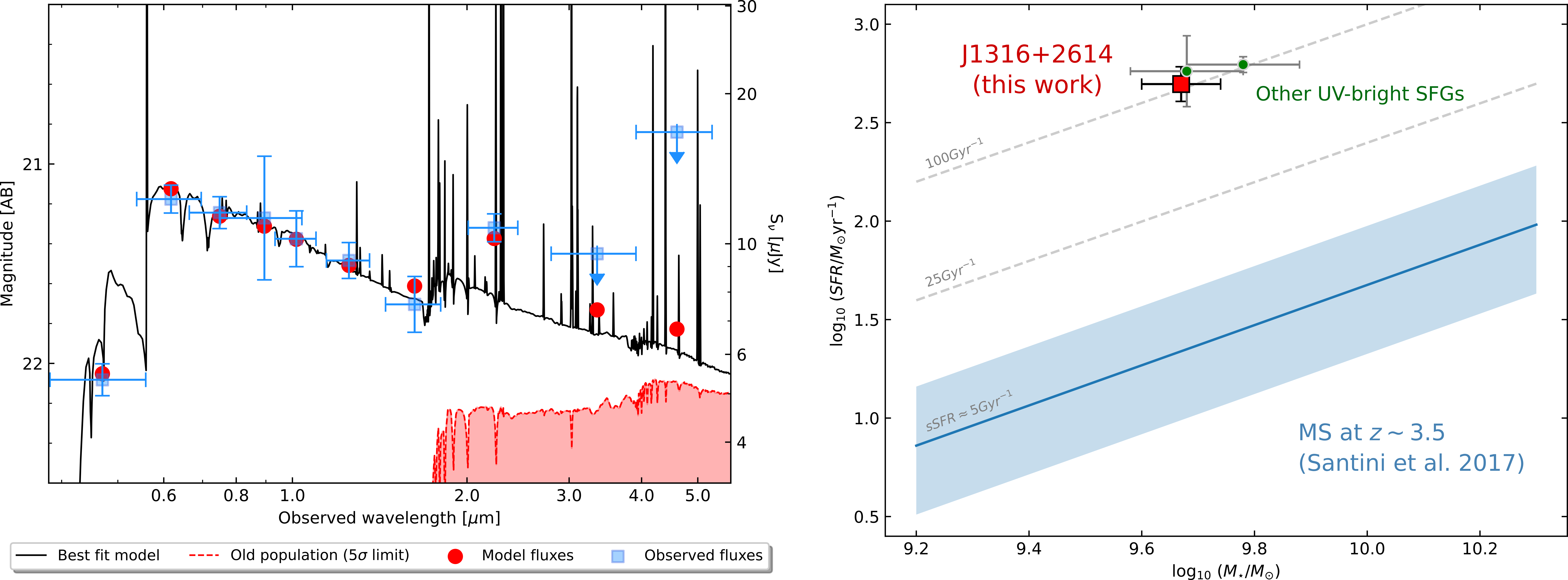}
  \caption{Left: SED best-fit model (black) of J1316$+$2614 using CIGALE \citep{Burgarella2005}. The fit uses photometry from $g$ to WISE $W2$ (blue squares, see Table~\ref{table1}), covering the rest-frame wavelength $0.1 - 1.0\mu$m. The predicted fluxes from the best-fit in each band are marked with red circles. The SED of J1316$+$2614 is dominated by a young stellar population with an age $14 \pm 4$~Myr and a continuous star formation rate  $SFR = 496 \pm 92 M_{\odot}$~yr$^{-1}$. The young stellar population has a stellar mass  $\log (M_{\star}/M_{\odot})=9.67 \pm 0.07$ with a dust attenuation $E(B-V)=0.035\pm0.016$. The old stellar population is not well constrained and we infer a $5\sigma$ limit for the stellar mass $\log (M_{\star}/M_{\odot}) \leq 10.3$ for a 580~Myr age (red shaded region). Right: Relation between $SFR$ and stellar mass of J1316$+$2614 (red square) and the other few star-forming galaxies known brighter than $M_{\rm UV} < -24$ (green circles, from \citealt{marques2020b} and \citealt{marques2021}, and assuming a \citealt{chabrier2003} IMF). The solid blue line and region mark respectively the observed main-sequence and scatter of star-forming galaxies at $z\simeq 3.5$ of \protect\cite{santini2017}.}
  \label{fig4}
\end{figure*}

Figure \ref{fig4} (left) shows the inferred photometry of J1316$+$2614. It shows a blue SED with optical to mid-IR color $r-W1 < -0.3$. This is much bluer than the typical colors measured in type-I/II AGNs at similar redshift, $r-W1 \simeq 0.6$ \citep{selsing2016, paris2018}. This is another evidence of the lack of a strong/dominant AGN component in J1316$+$2614.

To investigate the SED properties of J1316$+$2614, we perform SED-fitting with CIGALE code \citep{Burgarella2005, Boquien2019} using the available photometry covering a rest-frame wavelength $0.1 - 1.0\mu$m. The fit also includes flux measurements of the H$\beta$+[O~{\sc iii}] emission lines (Table \ref{table1}). 
We assume a star-formation history (SFH) with two components, a young stellar component with a continuous SFR with age $\leq 20$~Myr (see Section \ref{section31}), and an exponentially declining SFH with age of $\geq 200$~Myr to probe the old stellar population.
Stellar population models from \cite{bruzual2003}, \cite{chabrier2003} IMF, and the \cite{calzetti2000} dust attenuation law are considered. 
We fix the metallicity $Z/Z_{\odot}=0.4$ based on our analysis in Section~\ref{section31}, and the escape of ionizing photons is set as a free parameter (i.e., $0.0-1.0$). 

The best-fit model obtained for J1316$+$2614 is shown in the left panel of Figure \ref{fig4} (black). 
The young stellar population is characterized by an age of $14 \pm 4$~Myr and $\rm SFR = 496 \pm 92$~$M_{\odot}$~yr$^{-1}$, consistent with the far-UV analysis in Section~\ref{section31}. We infer a burst mass $\log (M_{\star}/M_{\odot})=9.67 \pm 0.07$ and dust attenuation $E(B-V)=0.035\pm0.016$. The LyC escape fraction is $f_{\rm esc} \rm (LyC) = 0.65 \pm 0.09$.
On the other hand, the mass of the old stellar population (age $580\pm285$~Myr) is not well constrained and we provide a $5\sigma$ limit of $\log (M_{\star}/M_{\odot}) \leq 10.3$ (in red in  Figure~\ref{fig4}, left). 
The SED properties derived from the multi-wavelength photometry agree well with those inferred from the UV fitting.

Our results indicate that the rest-frame UV to near-IR SED of J1316$+$2614 is dominated by the young stellar population, and if an old stellar population is present it should be less massive than $\log (M_{\star}/M_{\odot}) \leq 10.3$ (5$\sigma$). Considering the stellar mass of the young stellar population, we infer a high specific SFR (sSFR=SFR/$M_{\star}$) of $105\pm49$~Gyr$^{-1}$. The right panel of Figure \ref{fig4} shows the position of J1316$+$2614 in the SFR vs. $M_{\star}$ diagram and a comparison with main-sequence (MS) star-forming galaxies at similar redshifts \citep{santini2017}. J1316$+$2614 is offset of the MS by 1.5~dex. The other few star-forming galaxies known brighter than $M_{\rm UV} < -24$ at $z \sim 3$ also show similar $\rm sSFR \sim 100$~Gyr$^{-1}$ (green in Figure \ref{fig4}, \citealt{marques2020b, marques2021}). Even considering the mass of the old stellar population, J1316$+$2614 ($M_{\star}^{\rm total}/M_{\odot} \leq 10.6$) will be still located above the MS by $\simeq 0.7$~dex.

\section{Discussion}\label{section4}

J1316$+$2614 is one of the most luminous star-forming galaxies in the UV and Ly$\alpha$ discovered so far, with $M_{\rm UV}=-24.68\pm0.08$ and log$_{10}$(Ly$\alpha$\,/ erg~s$^{-1}$) = $44.08\pm0.10$, and in addition, one of the strongest LyC emitters known. In this section, we discuss its properties, compare with those of other strong LyC emitters, and discuss the implications of such discovery. Table \ref{table4} summarizes the main properties of J1316$+$2614.

\begin{table}
\begin{center}
\caption{Summary of the properties of J1316$+$2614. \label{table4}}
\begin{tabular}{l c c}
\hline \hline
\smallskip
\smallskip
  & Value & Uncertainty  \\
\hline 
R.A. (J2000)  & 13:16:29.61 & $0.1^{\prime \prime}$ \\
Dec. (J2000)  & $+$26:14:07.05 & $0.1^{\prime \prime}$ \\
$z_{\rm sys}$   & 3.6130 & $0.0008$   \\
$M_{\rm UV}$ (AB)   & $-24.68$ & $0.08$ \\
$\beta_{\rm UV}$  & $-2.59$  & $0.05$ \\
log(L[Ly$\alpha$ / erg~s$^{-1}$]) & 44.09 & $0.10$  \\
log(L[H$\beta$ / erg~s$^{-1}$]) & 43.18 & $0.10$  \\
$EW_{0} \rm (Ly\alpha)$ (\AA) & 20.5 & 1.9 \\
$EW_{0} \rm (H\beta)$ (\AA) & 34.7 & 6.8 \\
$Z_{\star}/Z_{\odot}$     & 0.4   &   [0.2 - 1.0] \\
$12+ \log$(O/H) & 8.45 & 0.12 \\
Age (Myr) [young]  & $9-14^{a}$   &   $[5-20]^{a}$  \\
log($M_{\star}/M_{\odot}$) [young]   & 9.67 & $0.07$ \\
log($M_{\star}/M_{\odot}$) [old]   & $\leq 10.3$ & --- \\
E(B-V)  & $0.006 - 0.034^{a}$   &   $[0 - 0.05]^{a}$   \\
SFR ($M_{\odot}$~yr$^{-1}$) & 497 & 92   \\
sSFR (Gyr$^{-1}$)  & 105   & 49  \\ 
$\Sigma$SFR ($M_{\odot}$~yr$^{-1}$~kpc$^{-2}$) & $\geq 10$ & ---  \\
log($\xi_{\rm ion}[\rm Hz~erg^{-1}]$) & $25.40$ & $0.10$ \\
$f_{\rm esc}$ (Ly$\alpha$) & 0.43 & 0.12 \\
$f_{\rm esc, rel}$ (LyC) & 0.92 & [$-0.20$, $+0.08$]  \\
$f_{\rm esc, abs}$ (LyC) &  0.87 & [$-0.20$, $+0.08$]  \\
log($Q^{\rm esc}_{H}$ / s$^{-1}$) & 55.86 & 0.11 \\
\hline 
\end{tabular}
\end{center}
\textbf{Notes. ---} (a) Age of the young stellar population and dust attenuation obtained with different methods: using UV spectral features and from the best-fit model of the SED using CIGALE, respectively. 
\end{table}

\subsection{AGN contamination or low-z interloper?}\label{section41}

We start investigating the possible contribution of an AGN or a low redshift contaminant to the flux detected below $\lambda_{0} < 912$\AA. 

\begin{figure*}
  \centering
  \includegraphics[width=0.80\textwidth]{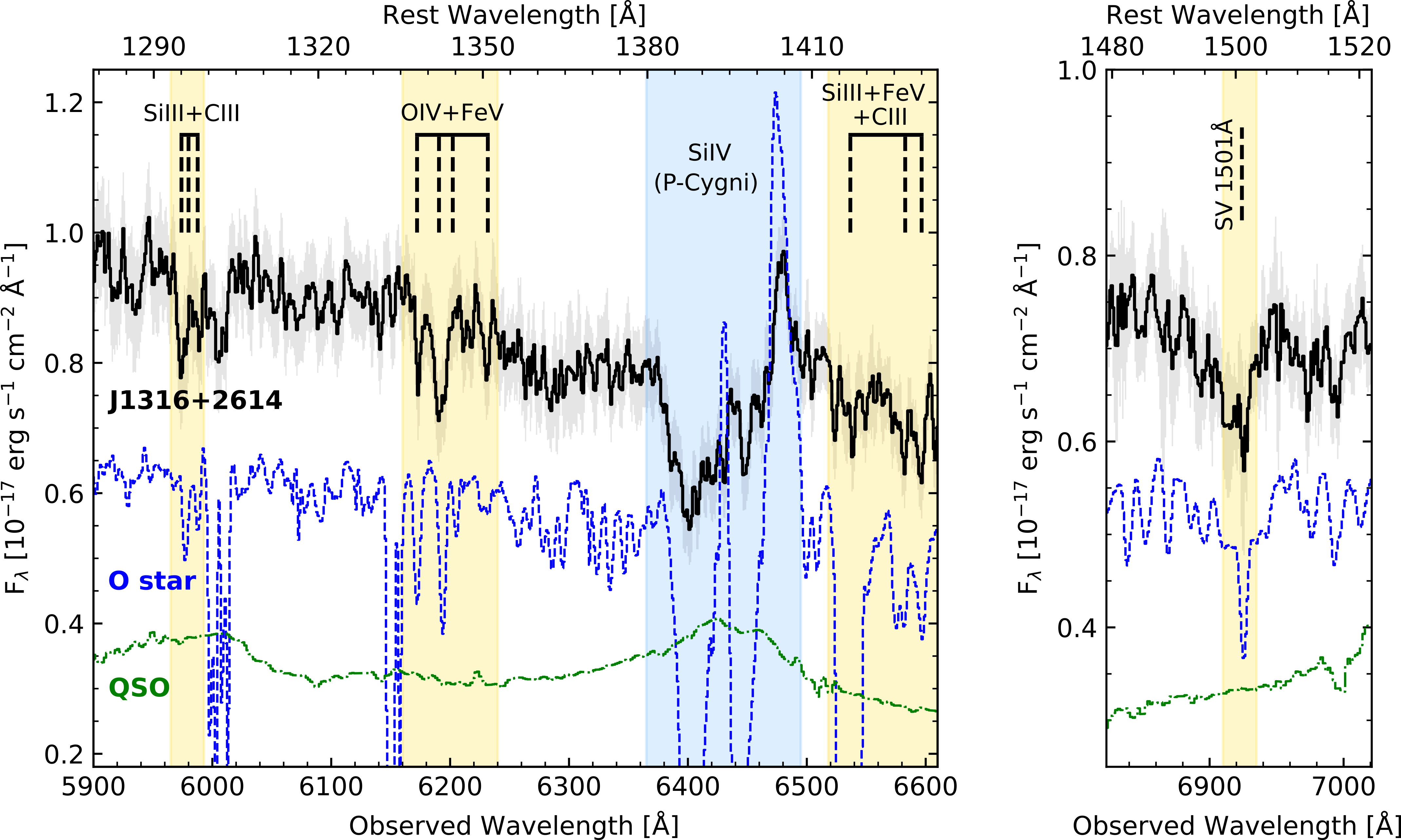}
  \caption{
  Zoom-in of the GTC spectrum (black, smoothed for visual propose) around $\lambda_{0} \simeq 1300-1400$\AA{ }(left) and $\lambda_{0} \simeq 1500$\AA{ }(right), where several photospheric lines and a P-Cygni profile in Si~{\sc iv} are detected (highlighted in yellow and blue shaded regions). For comparison, the HST/STIS spectrum of a O8.5 Iaf type star from the Large Magellanic Cloud (LH 9-34, \citealt{roman2020}) and the XShooter composite spectrum of bright QSOs \citep{selsing2016} are also shown in blue and green, respectively. }
  \label{fig55}
\end{figure*}

The rest-frame UV spectrum of J1316$+$2614 shows a wealth of absorption lines that are originated in the photospheres of hot stars. Figure~\ref{fig55} shows a zoom-in of the GTC spectrum around $\lambda_{0} \simeq 1300-1400$\AA{ }and $\lambda_{0} \simeq 1500$\AA, where some of these lines are clearly detected. 
The detection of these intrinsically weak photospheric lines indicates that the UV continuum of J1316$+$2614 is dominated by stellar emission, rather than an AGN, otherwise these absorption lines would be washed-out. For illustration, Figure \ref{fig55} also shows the spectrum of a O8.5 Iaf type star from the Large Magellanic Cloud observed with HST/STIS (LH 9-34, obtained as part of the ULLYSES program \citealt{roman2020}). In addition we also show the composite spectrum of bright QSOs of \cite{selsing2016} obtained with X-Shooter. As clearly seen in the figure, photospheric lines (highlighted in yellow) are  detected both in the spectra of J1316$+$2614 and the LH 9-34 star, but are not present in the spectra of QSOs.

Other spectral features, such as the P-Cygni profiles in wind lines of N~{\sc v}~1240\AA{ }and C~{\sc iv}~1550\AA, can be well explained by a young stellar population (Section~\ref{section31}), that is consistent with the multi-wavelength SED (Section~\ref{section34}). J1316$+$2614 shows a blue color between UV and near-IR (rest-frame) of $r-W1 < -0.3$, that is difficult to explain by a type-I/II AGN component. AGNs show typically much redder colors ($r-W1 \simeq 0.6$ \citealt{paris2018}) than the observed ones. J1316$+$2614 also shows a very blue UV slope $\beta_{\rm UV} = -2.60$, much steeper than those commonly observed in QSOs ($\simeq -1.0$ to $\simeq -1.7$, \citealt{lusso2015, selsing2016, liu2022}, but see \citealt{lin2022}). Lastly, the nebular emission in J1316$+$2614 presents narrow profiles without any hint for a broad component (FWHM > 1000~km~s$^{-1}$).

Our results clearly indicate that the large luminosity of J1316$+$2614 is being powered by massive star-formation. These do not discard the presence of a low luminosity AGN (e.g., Seyfert 2 or LINER), but rather indicate that a possible contribution of an AGN to the UV emission should be minimal. The same should be applied to the LyC emission. To quantify this, we consider conservatively a significant contribution of an AGN to the UV continuum of J1316$+$2614 of $\sim 25\%$, which is compatible with an AGN with an absolute magnitude $M_{\rm UV}^{\rm AGN} \simeq -23.2$. Assuming that the observed LyC flux arises from the AGN, this yields  $(f_{\rm LyC} / f_{1500})^{\rm obs}_{\rm AGN} \simeq 0.60$. AGNs with LyC detection at similar redshift and luminosities show much lower values ($(f_{\rm LyC} / f_{1500})_{\rm AGN} \simeq 0.05-0.25$; \citealt{lusso2015, cristiani2016, micheva2017}), even noting that the $f_{\rm esc} \rm (LyC)$ in these AGNs can be as high as 100\%. Therefore, a large contribution of an AGN to the LyC emission of J1316$+$2614 is highly unlikely.

Another source of contamination to the observed LyC emission could arise from a low-$z$ interloper. J1316$+$2614 appears unresolved in optical and near-IR images, but the poor spatial resolution from these observations ($\simeq 0.9^{\prime \prime}-1.2^{\prime \prime}$ FWHM) prevents us to rule out the presence of a low-$z$ interloper close to J1316$+$2614. From the low-resolution 2D spectrum, the spatial profiles of the emission observed below and above the Lyman edge at $\lambda_{\rm obs} \simeq 4200$\AA{ }(i.e., $\lambda_{\rm rest} \simeq 912$\AA) are similar, both unresolved and co-spatial. This means that, if a low-$z$ interloper is present, it should be co-spatial with J1316$+$2614. 

Considering the presence of a contaminant source at $z \gtrsim 0.5$, the emission seen $\lambda_{\rm obs} < 4200$\AA{ }would correspond to rest-frame UV radiation and, therefore, strong emission would be expected from the main rest-frame optical lines (e.g., [O~{\sc ii}], [O~{\sc iii}], H$\beta$ or H$\alpha$). However, we do not detect any other emission line within the spectral range covered by our optical and near-IR spectra (0.36$\mu$m to 1.00$\mu$m with GTC/OSIRIS and SDSS, and 1.45$\mu$m to 2.41$\mu$m with GTC/EMIR), rather than those originated from J1316$+$2614 at $z=3.613$. A high redshift contaminant ($z \gtrsim 2$) is also unlikely, because the flux density measured below $\lambda_{\rm obs} \lesssim 4200$\AA{ }is already very large ($\simeq 1.70\mu$Jy or $m \simeq 23.3$~AB). Lyman break galaxies or Ly$\alpha$ emitters at $z \geq 2$ are much fainter in the UV with typical aparent magnitudes at $\lambda_{\rm rest}$ of ($m_{\rm UV} \simeq 24.5-25.5$, e.g., \citealt{reddy2009}). Finally, the GTC spectrum of J1316$+$2614 shows absorption features in the LyC spectral range, that are consistent with Lyman series at redshifts between $z\simeq 3.034$ and $z\simeq 3.331$ (systems \#1, \#2, and \#3 in Figure~\ref{fig3}). This suggests that the emission observed at $\lambda_{\rm obs} < 4200$\AA{ }arises from the LyC emission of J1316$+$2614.

\subsection{Intrinsically high LyC leakage from indirect tracers}\label{section42}

In the previous section we shown that an AGN or a lower-$z$ interloper should not contribute significantly to the flux observed below $\simeq 4200$\AA{ }(i.e., the Lyman limit at $z=3.613$). Here we discuss the evidence towards a high LyC leakage in J1316$+$2614 using indirect methods (summarized in Table~\ref{table5}).

\begin{table}
\begin{center}
\caption{Direct and indirect LyC measurements of J1316$+$2614. \label{table5}}
\begin{tabular}{l c c c}
\hline \hline
\smallskip
\smallskip
Method & $f_{\rm esc, abs}$~(LyC) &  $\sigma$ & Section\\
\hline 
Direct LyC obs. & 0.87 & $[-0.20,+0.08]$ & \ref{section33}  \\
Multi-wav. SED & 0.65 & $\pm 0.09$ & \ref{section34} \\
$\beta_{\rm UV}$ vs. $EW_{0}~(H\beta)$ & $\approx 0.8$ & $>0.5$ & \ref{section421}  \\
$L(H\beta)$ & 0.82 & $\pm 0.1$ & \ref{section4211}  \\
ISM abs. lines & >$0.65$ & $3\sigma$ & \ref{section422}  \\
$\Delta v \rm (Ly\alpha$) & 0 & --- & \ref{section423}  \\
\hline 
\end{tabular}
\end{center}
\end{table}

\subsubsection{Weak nebular lines and continuum emission: $\beta_{\rm UV}$ and $EW_{0}\rm (H\beta)$ plane}\label{section421}

One evidence for high leakage is the \textit{weak} nebular lines and continuum emission in J1316$+$2614. For a young stellar population of $\simeq 9-14$~Myr old, strong nebular emission would be expected. Assuming a 9~Myr stellar population with a continuous star-formation history, an IMF with an upper mass limit of $100 M_{\odot}$, and $Z=0.4Z_{\odot}$, models predict equivalent widths in the Ly$\alpha$ and H$\beta$ recombination lines of $EW_{0}\rm (Ly\alpha)\simeq 120$\AA{ }and $EW_{0}\rm (H\beta) \simeq 145$\AA, respectively \citep{schaerer2009}. The predicted $EW_{0}$s are a factor of $\simeq 4-6\times$ larger than the ones observed in J1316$+$2614 (Table \ref{tab2}). The weak nebular emission also affects the SED of J1316$+$2614 (Figure~\ref{fig4}), making the observed fluxes in the $K_{\rm s}$ and WISE~3.6$\mu$m bands much weaker than expected if $f_{\rm esc}=0$, due to the contribution of H$\beta+$[O~{\sc iii}] and H$\alpha$, respectively.
In addition, the $\beta_{\rm UV} = -2.59 \pm 0.05$ measured in spectrum of J1316$+$2614 is similar to the intrinsic (best-fit) S99 stellar model, $\beta_{\rm UV} \rm (S99) = -2.62$, i.e., without considering the inclusion of nebular continuum or dust attenuation. This clearly suggests that the contribution of the nebular continuum and dust attenuation should be residual in J1316$+$2614. 

\begin{figure}
  \centering
  \includegraphics[width=0.47\textwidth]{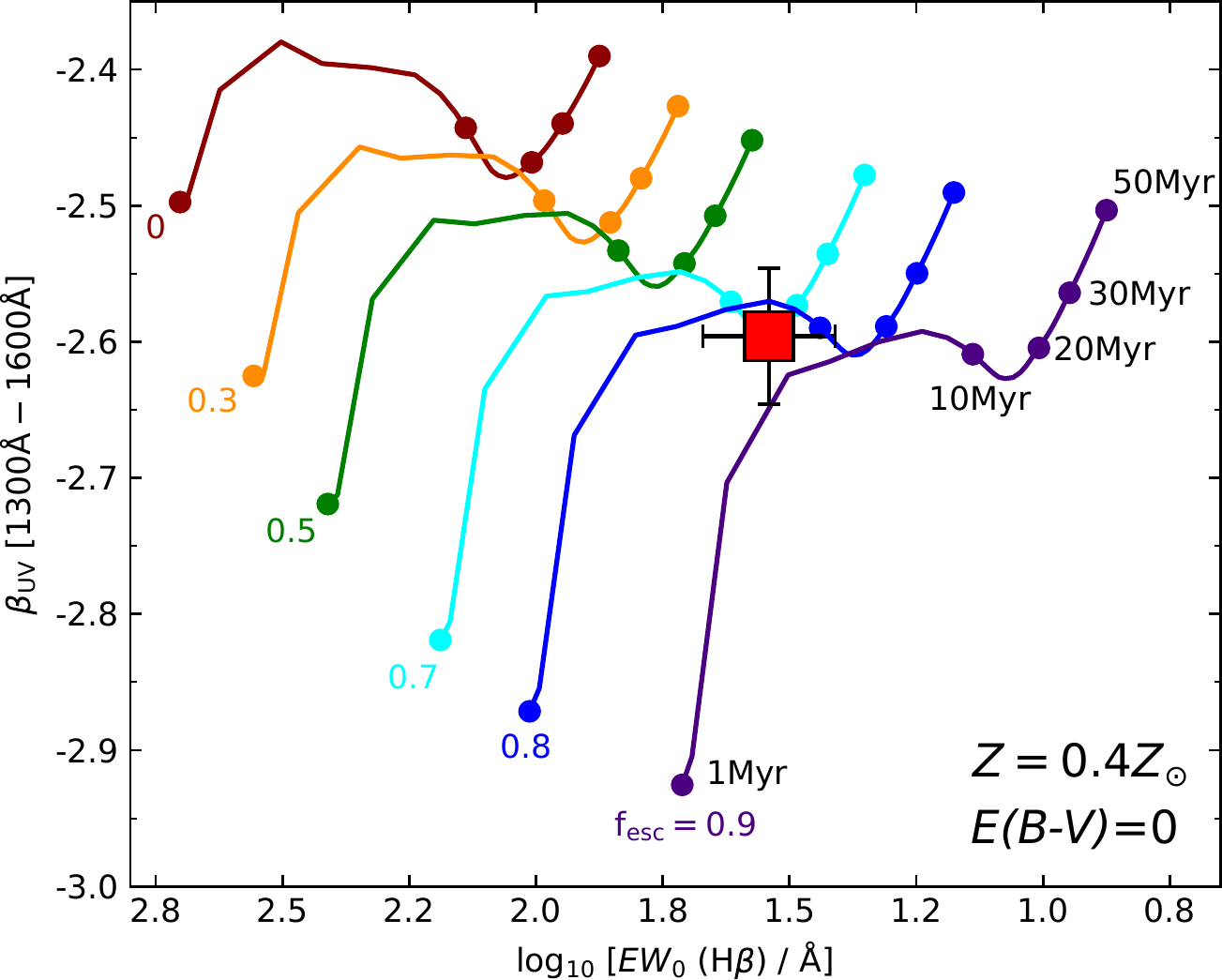}
  \caption{UV spectral slope ($\beta_{\rm UV}$) and H$\beta$ equivalent width ($EW_{0}$ (H$\beta$) as a function of $f_{\rm esc} \rm (LyC)$, predicted by S99 models with a continuous star-formation history with ages between 1$-$50 Myr and $Z = 0.4 Z_{\odot}$. For the H$\beta$ emission we use the models of \protect\cite{schaerer2009} matched to the same stellar properties. The position of J1316$+$2614, marked with a red square, suggests large $f_{\rm esc} \rm (LyC) \simeq 0.8$ at the age of 9 Myr, and discards $f_{\rm esc} \rm (LyC) \leq 0.5$. }
  \label{fig5}
\end{figure}

Since the escape of ionizing photons into the IGM/CGM should not contribute to photoionization in the ISM, a large LyC leakage would explain the weak nebular contributions in J1316$+$2614. To demonstrate this, Figure \ref{fig5} shows the relation between $\beta_{\rm UV}$ and $EW_{0}\rm (H\beta)$ as a function of $f_{\rm esc}$~(LyC) following the predictions of \cite{zackrisson2013} and \cite{zackrisson2017}. For this, we assume S99 models with a continuous star-formation history with ages between 1-50~Myr, and a metallicity $Z=0.4Z_{\odot}$. For H$\beta$, we use the models of \cite{schaerer2009} matched to the same stellar properties. Nebular continuum and emission are reduced by a factor of ($1-f_{\rm esc}$(LyC)). No dust attenuation is considered, since this effect is expected to be residual in J1316$+$2614.
As seen in this figure, the observed $\beta_{\rm UV}$ and $EW_{0}\rm (H\beta)$ in J1316$+$2614 (red square) is well explained by an $f_{\rm esc} \rm (LyC) \simeq 0.8$ at the age of 9~Myr. The figure also shows that $f_{\rm esc} \rm (LyC) \leq 0.5$ can be ruled out. This is compatible with the $f_{\rm esc} \rm (LyC) \simeq 0.87$ inferred in Section \ref{section33} using direct observations of the LyC emission. 

To our knowledge, this is the first observational confirmation that large escape of ionizing photons strongly affects the strength of nebular lines and continuum emission in a star-forming galaxy, thereby its global SED, as predicted by \cite{zackrisson2013}. For the few LyC emitters known with $f_{\rm esc} \rm (LyC) \gtrsim 0.5$ \citep{debarros2016, izotov2018b, flury2022b}, all of them show a combination of $EW_{0}\rm (H\beta)$ ($>100$\AA) and $\beta_{\rm UV}$ that, according to the  $\beta_{\rm UV}$ and $EW_{0}\rm (H\beta)$ plane, are incompatible with the inferred $f_{\rm esc} \rm (LyC)$ (e.g., see Figure 21 in \citealt{flury2022b}). An exception could be Ion2 \citep{vanzella2016} that shows a very steep UV slope $\beta_{\rm UV} = -2.7 \pm 0.6$ \citep{debarros2016} and $EW_{0}\rm (H\beta) \simeq 100$\AA{ }\citep{vanzella2020}, that according to Figure~\ref{fig7} is compatible with very high $f_{\rm esc} \rm (LyC)$, although with large uncertainties. It is known that the predicted tracks of $f_{\rm esc} \rm (LyC)$ in the $\beta_{\rm UV}$ and $EW_{0}\rm (H\beta)$ plane can be affected by different sources of uncertainty, such as the effect of dust, stellar properties, star-formation histories \citep{zackrisson2017}, and ultimately by an inhomogeneous LyC leakage. For J1316$+$2614, these appear well constrained. 

Finally, noting that J1316$+$2614 remains the only source known showing this behaviour (but see also J0121+0025 with $f_{\rm esc} \rm (LyC) \simeq 0.4$ and $EW_{0}\rm (Ly\alpha) \simeq 14$\AA; \citealt{marques2021}, or Ion1 with no Ly$\alpha$ emission at all; \citealt{ji2020}), our results highlight that at least some {\it weak} emission line galaxies with steep UV slopes could be potentially very strong LyC leakers. 
Our results do not contradict previous findings where $f_{\rm esc} \rm (LyC)$ is found to correlate with the strength of nebular emission lines (e.g., $EW_{0} \rm (Ly\alpha)$, $EW_{0} \rm (H\beta + [OIII])$, e.g., \citealt{izotov2016, flury2022b}), at least up to moderately high $f_{\rm esc} \rm (LyC)$. It rather indicates that the number of ionizing photos produced by the stellar population should be balanced with those contributing to the photoionization of H~{\sc ii} regions and those escaping the galaxy, which can be critical at very high $f_{\rm esc} \rm (LyC)$ ($\geq 0.5$).

\subsubsection{$H\beta$ Luminosity}\label{section4211}

Another straightforward way to test $f_{\rm esc} \rm (LyC)$ is to use the H$\beta$ luminosity. Since H$\beta$ is a recombination line and has little dependence on the metallicity and electron temperature ($T_{e}$) \citep[e.g.,][]{charlot2001}, it is basically a direct photons {\it counter}. Assuming $T_{e}=10^{4}K$, the H$\beta$ luminosity ($L\rm (H\beta)$) can expressed in terms of the production rate of ionizing photons, $Q^{\rm int}_{H}$, and $f_{\rm esc} \rm (LyC)$ by:

\begin{equation}
    \begin{aligned}
        L(H\beta)& = \frac{\gamma_{\rm H\beta} \: Q^{\rm int}_{\rm H}}{\alpha_{B}} \times (1-f_{esc})
        \\ \noalign{\vskip5pt}
        & = 2.1\times 10^{12} \: Q^{\rm int}_{\rm H} \times (1-f_{esc}),
    \end{aligned}
    \label{eq3}
\end{equation}

\noindent
where $\gamma_{\rm H\beta}$ is the nebular emission coefficient and $\alpha_{B}$ is the Case B recombination rate \citep{osterbrock2006}. 
From the $Q^{\rm int}_{\rm H} = 8.5 \times 10^{55}$~s$^{-1}$ inferred in Section \ref{section33} and $L\rm (H\beta)= (1.2\pm 0.3) \times 10^{43}$~erg~s$^{-1}$ derived from the observed H$\beta$ flux (Table \ref{tab2}), we get $f_{\rm esc} \rm (LyC) = 0.82$, that matches almost perfectly with the direct measurement of $f_{\rm esc} \rm (LyC)$.

\subsubsection{{\it Null} covering fraction of neutral gas}\label{section422}

Another piece of evidence for the high $f_{\rm esc} \rm (LyC)$ is the absence of ISM low-ionization absorption lines (LIS) in the spectrum of J1316$+$2614, that probe neutral gas in the line of sight \citep{gazagnes2018}. 

The spectra of typical LBGs or LAEs present typically relatively strong LIS lines, with $EW_{0} \rm (LIS) \sim (1-4)$\AA{ }\citep[e.g.,][]{shapley2003, shibuya2014, marques2020}, but these are very weak or not detected in J1316$+$2614, as well as in other strong LyC leakers \citep{rivera2017, chisholm2017b, chisholm2018, izotov2018b, marques2021, saldana2022}. The weakness of LIS lines in J1316$+$2614 could arise by a low geometric covering fraction of the gas, $C_{f}$, which would indicate a large $f_{\rm esc} \rm (LyC)$, as $C_{f}=1-f_{\rm esc} \rm (LyC)$ \citep[e.g.,][]{steidel2018}. $C_{f}$ can be inferred using the residual intensity of the absorption line, $I$, so that $C_{f} = 1 - I/I_{0}$, where $I_{0}$ is the continuum level.\footnote{This assumes the optically thick regime and an ionization-bounded ISM with a uniform dust-screen geometry.} Assuming a line width of $\rm FWHM = 200$~km~s$^{-1}$, we infer $I/I_{0} > 0.85$ and $EW_{0} \rm (LIS) < 0.6$\AA{ }at $3\sigma$ confidence level for Si~{\sc ii}~1260\AA{ }and C~{\sc ii}~1334\AA. A low geometric covering fraction of the gas is inferred for J1316$+$2614, $C_{f} < 0.15$ ($3\sigma$), that  would be consistent with a high $f_{\rm esc} \rm (LyC)$. Following the empirical correlation between the residual intensity of ISM LIS lines and $f_{\rm esc} (LyC)$ of \cite{saldana2022}, we infer $f_{\rm esc} \rm (LyC) > 0.65$ ($3\sigma$). A low ion column density could also explain the weakness of LIS lines, but such scenario is unlikely, because these lines are almost always saturated in the spectra of star-forming galaxies, even in damped-Ly$\alpha$ systems with very low metallicities \citep[e.g.,][]{des2006}.

\subsubsection{Ly$\alpha$ spectral profile}\label{section423}

Another common and well-studied tracer of $f_{\rm esc} \rm (LyC)$ is the spectral shape and peak separation ($\Delta v \rm (Ly\alpha$)) of the Ly$\alpha$ profile \citep[e.g.,][]{verhamme2015, verhamme2017, izotov2018b}, that depends on the amount and geometry of the neutral gas and dust. Following \cite{verhamme2015} and assuming a spherical homogeneous shell, the observed $\Delta v \rm (Ly\alpha)= 680 \pm 70$~km~s$^{-1}$ in J1316$+$2614 suggests a large H~{\sc i} column density $N_{\rm HI} \gtrsim 10^{20.5}$~cm$^{-2}$, implying $f_{\rm esc} \rm (LyC) \approx 0$ \citep{verhamme2017, izotov2018b}. A priori, this is incompatible with the $f_{\rm esc} \rm (LyC) \simeq 0.9$ directly measured from the LyC emission itself (Section \ref{section34}). This suggests that J1316$+$2614 may have an inhomogeneous ISM geometry (e.g., clumpy) and/or complex gas dynamics. 

With the available data it is not possible to draw any definitive conclusion on the connection between the Ly$\alpha$ profile and LyC leakage. The lack of low-ionization ISM absorption lines in the spectrum of J1316$+$2614 (Section \ref{section422}) indicates a low covering fraction of neutral gas in the line-of-sight. This is compatible with large $f_{\rm esc} \rm (LyC) \simeq 0.9$ observed in the spectrum of J1316$+$2614, and suggests a very low $N_{\rm HI}$ along the LyC source, $N_{\rm HI} \lesssim  10^{16}$~cm$^{-2}$ \citep{verhamme2017}. In such a case, Ly$\alpha$ would be preferentially seen at the systemic velocity (e.g., \citealt{naidu2022}), but it should be highly suppressed too given the large escape of ionizing photons (similarly as H$\beta$, see Section~\ref{section421}). The Ly$\alpha$ profile seen in the $R\simeq 1800$ OSIRIS spectrum (Figure \ref{fig2}) shows already $f_{\lambda}>0$ at the systemic redshift, but the emission is faint and the limited resolution prevents us to properly resolve the line. It is thus possible that Ly$\alpha$ is being emitted from regions spatially offset from the LyC source, and if so, it could be offset and possible more extended than the UV continuum. 

The shape of the observed Ly$\alpha$ profile and the properties of the neutral gas inferred from it also depends on the intrinsic nebular profile, which is not well characterized with our data. The non-resonant lines H$\beta$ and [O~{\sc iii}] have relatively low SNR and are observed with low resolution (Figure~\ref{fig4}), not enough for a proper study of the intrinsic emission (e.g., to look for broad wings or complex structure of the line).

While higher spectral and spatial resolution observations of Ly$\alpha$ and non-resonant lines are required to investigate better the complex Ly$\alpha$ profile and its connection to the LyC in J1316$+$2614, our results already indicate that large $f_{\rm esc} \rm (LyC)$ and large $\Delta v \rm (Ly\alpha$) can co-exist. It is worth noting that similar large $\Delta v \rm (Ly\alpha$) have been observed in a few sources with evidence of LyC leakage. For example, the local infrared galaxy IRAS~01003$-$2238 studied in \cite{martin2015} shows a remarkably similar Ly$\alpha$ profile as the one in J1316$+$2614, with a blue-to-red peak line ratio $I_{\rm blue}/I_{\rm red} > 1$ and similar peak separation. Complex kinematics with high velocity wings are found both in Ly$\alpha$ and in non-resonant lines (e.g., H$\beta$), suggesting that scattering merely enhances the wings relative to the line core. Using radiative transfer modeling and assuming a clumpy geometry, \cite{martin2015} found that the Ly$\alpha$ profile can be well explained by a low column density ($N_{\rm HI} \simeq 10^{17}$~cm$^{-2}$), that is optically thin to LyC. Another example is the LyC emitter Ion2 ($f_{\rm esc} \rm (LyC) \gtrsim 0.5$; \citealt{debarros2016, vanzella2016}), whose Ly$\alpha$ emission shows a peak separation of $\sim 550$~km~s$^{-1}$ when observed with relatively low resolution ($R\simeq 1200$, \citealt{debarros2016}). Higher resolution observations ($R\simeq 5300$, \citealt{vanzella2020}) reveal a much more complex Ly$\alpha$ profile, including a substantial emission at the systemic velocity.

Another important point worth mentioning is the difference between the Ly$\alpha$ and LyC escape fractions. While $f_{\rm esc}$~(Ly$\alpha) \geq f_{\rm esc} \rm (LyC)$ is usually assumed and predicted \citep{kimm2019, maji2022}, J1316$+$2614 shows 
$f_{\rm esc} \rm (Ly\alpha) = 0.43 \pm 0.12$ (Section~\ref{section33}), that is $\sim 2\times$ lower than $f_{\rm esc} \rm (LyC)$. However, we note that the Ly$\alpha$ emission represents only a small fraction ($\approx 5\%$) of the total LyC photons produced in J1316$+$2614, possibly indicating that the regions probed by Ly$\alpha$ could not be representative of those where LyC is escaping.

\subsubsection{Other properties}\label{section424}

Other observational signatures of J1316$+$2614 resemble those found in other LyC emitters. 
J1316$+$2614 shows a compact morphology, being not resolved in ground-based imaging. Assuming the $\simeq 1^{\prime \prime}$~FWHM seeing conditions of optical images, we estimate an effective radius in the rest-frame UV $r_{\rm eff} \leq 2$~kpc. Considering the large $\rm SFR \simeq 500$~$M_{\odot}$~yr$^{-1}$ inferred from the SED in Section \ref{section34}, this implies a large SFR surface density $\Sigma \rm SFR \geq 10$~$M_{\odot}$~yr$^{-1}$~kpc$^{-2}$, that is within the range of $\Sigma \rm SFR$ inferred in other strong LyC emitters \citep[e.g.,][]{vanzella2016, izotov2018b, marques2021, flury2022b}. Such an intense, concentrated star-formation and the expected feedback from stellar winds of massive stars and SN explosions could play a major role in shaping the ISM \citep[e.g,][]{sharma2017}, favouring the escape of ionizing photons. Although inflows appear to be the dominant kinematic pattern of the gas in J1316$+$2614, strong ionized outflows are also expected given the nature of J1316$+$2614 \citep[e.g.,][]{alvarez2021}. High SNR and high-resolution observations of the main rest-frame optical lines are required to investigate the presence of outflowing gas in J1316$+$2614. 

\begin{figure*}
  \centering
  \includegraphics[width=0.98\textwidth]{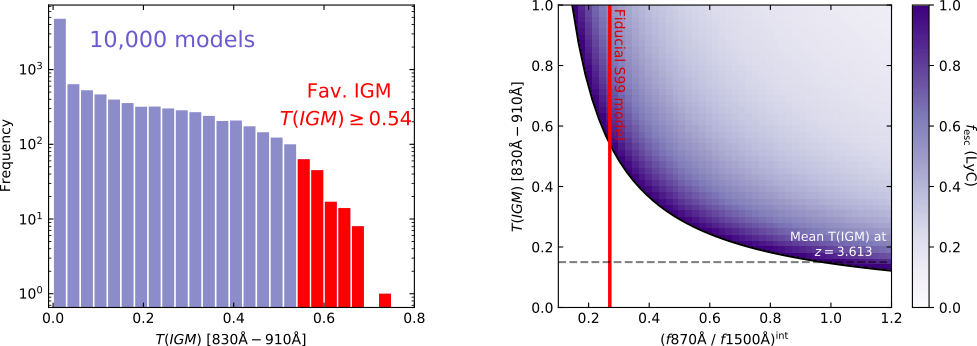}
  \caption{Left panel: frequency distribution of the IGM transmission ($T(IGM)$) of 10,000 model sightlines simulating the distribution of H~{\sc i} absorbers at $z=3.613$. These models were generated using TAOIST-MC code \citep{bassett2021} and $T(IGM)$ is computed between rest-frame wavelengths $830-910$\AA. Only a small fraction of these models (148/10,000) predicts $T(IGM) \geq 0.54$ that is needed to keep $f_{\rm esc} \leq 1$ in J1316$+$2614 (highlighted in red). Right panel: relationship between $T(IGM)$ and $(f_{870} / f_{1500})^{\rm int}_{\nu}$ as a function of $f_{\rm esc}$ (color bar) following Equation \ref{eq1}. Red vertical is the $(f_{870} / f_{1500})^{\rm int}$ inferred from the best-fit S99 model ($\simeq 0.27$). Horizontal dashed line is the mean $T(IGM)$ for the 10,000 models, $T(IGM) \simeq 0.15$. }
  \label{fig6}
\end{figure*}

On the other hand, J1316$+$2614 shows a relatively modest [O~{\sc iii}]/[O~{\sc ii}] line ratio, $O_{32}=4.8 \pm 2.1$, at least compared to other strong leakers that show $O_{32} \gtrsim 10$ \citep[e.g.,][]{debarros2016, izotov2018b}. Using the empirical relation of \cite{izotov2018a}, the observed $O_{32} \simeq 5$ predicts $f_{\rm esc} \rm (LyC) \simeq 0.05$, which is inconsistent with our observations. However, this relation has been constrained using green pea LyC leakers with very low metallicities (12+log(O/H)$\leq 8.0$). As shown by \cite{bassett2019}, line ratios of $O_{32} \lesssim 5$ can still predict large $f_{\rm esc} \rm (LyC) \gtrsim 0.5$ if higher metallicities and/or lower ionization parameters are considered, which seems to be the case of J1316$+$2614 (12+log(O/H)=$8.45\pm0.12$, see Section \ref{section32}). This is also consistent with the recent results of \cite{flury2022b}, where a large scatter in $O_{32}$ with $f_{\rm esc} \rm (LyC)$ is found in low-$z$ LyC emitters with a wide range of metallicities (12+log(O/H)=$7.5-8.6$).

Finally, the spectrum of J1316$+$2614 shows strong P-Cygni profiles in wind lines (N~{\sc v}~1240\AA, C~{\sc iv}~1550\AA, and also O~{\sc vi}~1033\AA, see Figures \ref{fig1} and \ref{fig3}) and these are ubiquitous in other strong LyC leakers \citep[e.g.,][]{borthakur2014, izotov2016, izotov2018a, izotov2018b, rivera2019, vanzella2018, vanzella2020, marques2021}. While the presence of these  spectral features does not necessarily imply LyC leakage, they indicate very young ages of the stellar population \citep[e.g.,][]{chisholm2019}, where the production of LyC photons from O-type stars and feedback from the strong winds and SN explosions are more efficient.

\subsection{Favorable IGM transmission and/or high LyC production efficiency?}\label{section43}

From the low-resolution OSIRIS spectrum we measure a ratio between the ionizing to non-ionizing flux density of $(f_{870} / f_{1500})_{\rm obs} = 0.146 \pm 0.011$. This is a factor of $\simeq 2-3$x higher than those measured in other strong LyC emitters at $z\simeq 3$ \citep[e.g.,][]{fletcher2019, pahl2021, saxena2022}, despite the fact that they are at lower redshifts and, thereby, they might be less affected by IGM absorption than J1316$+$2614. Following Equation \ref{eq1} and the intrinsic ratio of the best-fit S99 model, $(f_{870} / f_{1500})_{\nu}^{\rm int} = 0.27$, we find that the IGM transmission in the spectral region probed by the LyC emission at $\lambda_{0} = 830-910$\AA{ }should be $T(IGM) > 0.54$ to keep $f_{\rm esc} \leq 1$. This is significantly larger than the mean $T(IGM)$ expected at $z\simeq 3.5$ ($<T(IGM)>~\sim 0.2$, e.g.,  \citealt{inoue2014}, \citealt{steidel2018}, \citealt{bassett2021}). Here we investigate the probability to have such a favorable IGM transmission along the line of sight of J1316$+$2614.

We generate 10,000 model $z = 3.613$ sightlines simulating the distribution of H~{\sc i} absorbers using the {\sc TAOIST-MC} code\footnote{\url{http://github.com/robbassett/TAOIST_MC/}} \citep{bassett2021}. $T(IGM)$ functions use the models of \cite{inoue2014} and account for both intergalactic and circumgalactic medium (IGM+CGM) following the prescriptions of \cite{steidel2018}. For every model sightline, we derive the net transmission over the rest wavelength range $\lambda_{0} = 830-910$\AA, the same interval used to measure the LyC flux of J1316$+$2614. 

Left panel of Figure \ref{fig6} shows the $T(IGM)$ frequency distribution of these realizations. The bulk ($\simeq 80\%$) of the simulations have $T(IGM) < 0.4$. However, a small fraction of these models (148/10,000) shows a favorable IGM transmission (defined as $T(IGM) \geq 0.54$) that is compatible with $f_{\rm esc} \leq 1$. The mean IGM transmission at $830-910$\AA{ }for the 10,000 models is found to be $T(IGM)^{\rm mean} \simeq 0.15$, while for the favorable sightlines is $T(IGM)^{\rm fav} \simeq 0.58$, that is similar to that measured between $\lambda_{0} \simeq 950-1180$\AA{ }in Section \ref{section33} ($T(IGM) = 0.59 \pm 0.17$). Therefore, our results suggest that a favorable IGM transmission is still possible, although such occurrence has low probability ($P[T(IGM) \geq 0.54] \simeq 0.015$).

A relatively lower IGM transmission is still possible if the intrinsic ratio $(f_{870} / f_{1500})^{\rm int}$ is higher than the one assumed from the best-fit S99 model ($\simeq 0.27$), i.e., an SED with higher LyC production efficiency ($\xi_{\rm ion}$). 
The right panel of Figure \ref{fig6} shows the behaviour between $T(IGM)$ and $(f_{870} / f_{1500})^{\rm int}$ as a function of $f_{\rm esc}$~(LyC), following Equation \ref{eq1}. Considering the mean IGM transmission at $z = 3.613$, $T(IGM)^{\rm mean} \simeq 0.15$, an intrinsic $(f_{870} / f_{1500})^{\rm int} \geq 0.9$ is required to keep $f_{\rm esc} \leq 1$, i.e., a source producing at least $\geq 3\times$ more ionizing photons than our best-fit S99 model ($\xi_{\rm ion} \gtrsim 25.9$).

Such high values of $\xi_{\rm ion}$ are disfavoured for J1316$+$2614 as the age and metallicity of the young stellar population are relatively well constrained ($9\pm 5$~Myr and $Z/Z_{\odot} = 0.4$, Section \ref{section31}). Considering a younger age (4~Myr) and lower metallicity ($Z/Z_{\odot} = 0.2$), $(f_{870} / f_{1500})^{\rm int}$ would increase only a factor of $\simeq 20\%$ than our fiducial S99 model. Also, the inclusion of binaries can increase $(f_{870} / f_{1500})^{\rm int}$. Considering a BPASS models \citep[v2.2,][]{stanway2016} with a continuous star-formation history with the same age (10~Myr), metallicity ($Z/Z_{\odot} = 0.4$) and IMF, the increase of $(f_{870} / f_{1500})^{\rm int}$ is roughly around $\simeq 25\%$ when binaries are included. Finally, a different slope and a higher upper mass limit of the IMF can increase considerably $(f_{870} / f_{1500})^{\rm int}$. Considering the most top-heavy IMF model available in the BPASS library, with a IMF slope of $-2.0$ and an upper mass limit of $300M_{\odot}$, and a continuous star-formation history with 10~Myr and $0.4Z_{\odot}$, we infer a $(f_{870} / f_{1500})^{\rm int} \simeq 0.46$ ($\xi_{\rm ion} \simeq 25.60$). In such a case, a $T(IGM) \geq 0.3$ is needed to keep $f_{\rm esc} \leq 1$ (Figure \ref{fig6}, right), i.e., still a factor of $\simeq 2$x higher than the mean IGM transmission at $z=3.613$.

In short, our results suggests that, even considering a stellar population with a higher $(f_{870} / f_{1500})^{\rm int}$ and $\xi_{\rm ion}$ than that assumed for J1316$+$2614, a favorable IGM transmission and a large $f_{\rm esc}$~(LyC) are required to explain the LyC flux measured in this source. 

Finally, the previous discussion on $(f_{870} / f_{1500})^{\rm int}$, $T(IGM)$, and therefore $f_{\rm esc}$~(LyC) relies on the assumption that the observed LyC flux of J1316$+$2614 arises from the stellar emission itself, which might not be necessarily true. Models can predict the escape of free-bound emission of hydrogen in ionized nebulae from the radiation energy re-distribution of stellar LyC \citep[][]{inoue2010}. Under certain conditions (e.g., high electron temperature and $f_{\rm esc}$~(LyC), hard stellar SED, see details in \citealt{inoue2010}), this emission can be strong, producing a "Lyman bump" seen just below of the Lyman edge ($\lesssim 912$\AA). If present in J1316$+$2614, the $f_{\rm esc}$~(LyC) inferred in Section \ref{section31} could be overestimated, although it should be high. The limited SNR and spectral coverage of our OSIRIS spectrum in the LyC region, along with the inherent uncertainties due to IGM, prevent us to further investigate the presence of this "Lyman bump" emission in J1316$+$2614. Nevertheless, its presence could explain the absence of the Lyman break around $\simeq 912$\AA{ }in the spectrum of J1316$+$2614 (see Figure~\ref{fig3}), that is expected if only pure stellar models are assumed.

\subsection{The most powerful ionizing source known among star-forming galaxies}\label{section44}

\begin{figure*}
  \centering
  \includegraphics[width=0.95\textwidth]{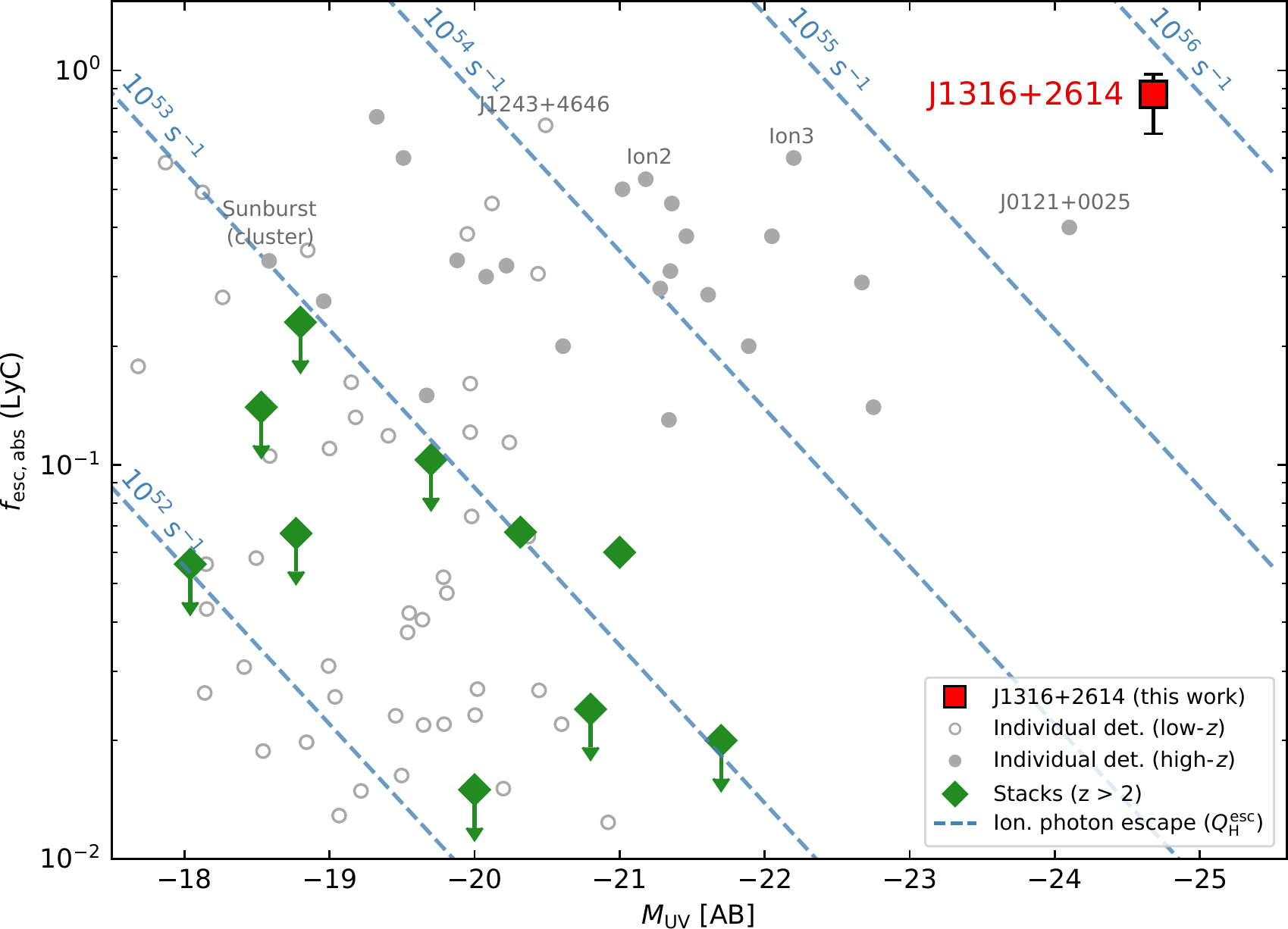}
  \caption{Relation between the absolute LyC escape fraction, $f_{\rm esc}$~(LyC), and UV absolute magnitude, $M_{\rm UV}$ (AB). J1316$+$2614 is represented with a red square. Other $\sim 40$ individual star-forming galaxies with direct detection of LyC are also shown. These include $z\sim0.3$ sources (\citealt{izotov2016, izotov2016b, izotov2018a, izotov2018b, flury2022a}, open circles) and other star-forming galaxies at $z>2$ (\citealt{vanzella2016, shapley2016, vanzella2018, fletcher2019, rivera2019, vanzella2020, marques2021, saxena2022}, solid circles). Results from deep stacks are marked with green diamonds (\citealt{grazian2017, marchi2017, rutkowski2017, fletcher2019, bian2020, pahl2021, begley2022}). Blue dahsed lines show the ionizing photon escape, $Q_{H}^{\rm esc}$, predicted for a given $M_{\rm UV}$ and $f_{\rm esc}$~(LyC) assuming the ionizing photon production efficiency derived for J1316$+$2614 (log($\xi_{\rm ion}/\rm erg^{-1}Hz) = 25.40$). J1316$+$2614 is by far the most powerful ionizing source known among star-forming galaxies, both in terms of $f_{\rm esc}$~(LyC) and $Q_{H}^{\rm esc}$.}
  \label{fig7}
\end{figure*}

We now proceed to compare the $f_{\rm esc}$~(LyC) and the UV absolute magnitude of J1316$+$2614 with other LyC emitters. Figure~\ref{fig7} shows the $f_{\rm esc}$~(LyC) and $M_{\rm UV}$ of J1316$+$2614 (magenta) and other $\sim 40$ individual LyC emitters with direct detection of LyC radiation. The comparison sample consists of both low redshift ($z\sim 0.3$, \citealt{izotov2016, izotov2016b, izotov2018a, izotov2018b, flury2022b}, open circles) and $z > 2$ star-forming galaxies (\citealt{vanzella2016, shapley2016, vanzella2018, fletcher2019, vanzella2020, marques2021, saxena2022}, solid circles). The figure also includes several statistical results of $f_{\rm esc}$~(LyC) from deep imaging and spectral stacks of LBGs and LAEs (\citealt{grazian2017, marchi2017, rutkowski2017, fletcher2019, bian2020, pahl2021, begley2022}, green diamonds). Most of these stacks show upper limits on $f_{\rm esc}$~(LyC) from $\sim 0.2$ to a few percent.

As seen in this figure, the inferred $f_{\rm esc}~\rm (LyC) \simeq 0.87$ of J1316$+$2614 places it among the most powerful LyC emitter known. This result is robust and should not depend significantly with the assumptions on the IGM attenuation. Even assuming a completely transparent (and unrealistic) IGM for J1316$+$2614, $T(IGM)=1$, we would still infer $f_{\rm esc} \rm (LyC) \simeq 0.54$ (Section~\ref{section33}), that is larger than in many other LyC emitters. Therefore, our results indicate that large $f_{\rm esc}$~(LyC) can be found in UV-bright star-forming galaxies, and are not restricted to the faintest ones as previously thought, as already discussed in \cite{marques2021}. Previous studies have found a possible anti-correlation between $f_{\rm esc}$~(LyC) and $M_{\rm UV}$ \citealt{steidel2018, pahl2021, saldana2022}, i.e., $f_{\rm esc}$~(LyC) decreases towards UV-brighter sources, but these works probe relatively faint galaxies with a narrow range of UV absolute magnitudes ($M_{\rm UV} \sim -20 \pm 1$). To our knowledge there are no statistical studies probing $f_{\rm esc}$~(LyC) of star-forming galaxies brighter than $M_{\rm UV} < -22$. The few exceptions are the ones presented in \cite{grazian2017} where $f_{\rm esc} \rm (LyC) < 0.08$ ($3\sigma$) is found for $M_{\rm UV} \sim -23$, but their sample consists of two sources only. Figure~\ref{fig7} clearly shows that several UV-bright sources ($M_{\rm UV} < -22$, \citealt{vanzella2018, marques2021, saxena2022}), including J1316$+$2614, are very strong LyC emitters. 

It is worth discussing that many properties inferred for J1316$+$2614 and in other UV-bright LyC emitters (\citealt{vanzella2018, marques2021}) are remarkably similar as those found in UV-faint leakers (e.g., \citealt{izotov2018b}). These include very young ages ($\lesssim 10$~Myr), high $sSFR$ ($\sim 100$~Gyr$^{-1}$), compact morphologies and low dust attenuation, among others (e.g., weak UV absorption lines, strong wind lines, etc.). The remarkably difference between them is the $\simeq 4-5$ magnitudes in the UV, that is related with the strength or amplitude of the burst. The Ly$\alpha$ profile of J1316$+$2614 also differs from those observed in other LyC leakers, showing typically a red-dominated peak and $\Delta v \rm (Ly\alpha) \lesssim 250$~km~s$^{-1}$  \citep[e.g.,][]{izotov2018b}. It is still unclear if the properties governing the escape of ionizing photons of these UV-bright galaxies are similar to those of UV-faint sources, but should be further explored with larger samples of UV-bright sources. 

Another important output from Figure~\ref{fig7} is the ionizing photon escape ($Q_{H}^{\rm esc}$) predicted for a given $M_{\rm UV}$ and $f_{\rm esc}$~(LyC). This is highlighted by the dashed lines in Figure~\ref{fig7}. For simplicity, these lines assume the ionizing photon production efficiency derived for J1316$+$2614, log($\xi_{\rm ion}/\rm erg^{-1}Hz) = 25.40$. We caution that the assumed $\xi_{\rm ion}$ may differ from source to source depending on galaxy properties (e.g., metallicity, age, SFH). For example, some UV-faint, low-metallicity starbursts at low-$z$, including LyC emitters, present higher $\xi_{\rm ion}$ (e.g., log($\xi_{\rm ion}/\rm erg^{-1}Hz) \simeq  25.8$, \citealt{schaerer2016}), while more typical $z\sim 2-5$ LBGs and LAEs show log($\xi_{\rm ion}/\rm erg^{-1}Hz) \simeq  24.8-25.4$ (e.g., \citealt{bouwens2016b, matthee2017b}). Despite the differences in the assumed $\xi_{\rm ion}$, Figure~\ref{fig7} shows that the ionizing photon escape of J1316$+$2614 ($Q_{H}^{\rm esc} \approx (7-8)\times 10^{55}$~s$^{-1}$, Section~\ref{section34}) is orders of magnitude higher than those predicted in UV-faint sources, even those showing large $f_{\rm esc}$~(LyC). The differences are even more extreme if the results from deep stacks of typical LBGs and LAEs are considered, where LyC emission is usually not detected down to $f_{\rm esc}$~(LyC) of a few per cent \citep{rutkowski2017, fletcher2019}. The $Q_{H}^{\rm esc}$ inferred for J1316$+$2614 is $\geq 10^{3} - 10^{4}$ higher than that inferred for stacks of LAE/LBG population.

The combination of large production and escape of LyC photons makes J1316$+$2614 the most powerful ionizing source known among star-forming galaxies, only comparable to J0121$+$0025 at $z=3.24$ ($M_{\rm UV} = -24.2$ and $f_{\rm esc} \rm (LyC) \approx 0.4$; \citealt{marques2021}). Whether or not these UV-luminous star-forming galaxies can contribute to the cosmic reionziation at higher redshifts ($z \gtrsim 6$), locally or globally, is still unclear and depends fundamentally if such sources are present in the early Universe and on their number density. Recent works have found exceptionally luminous ($M_{\rm UV} \lesssim -23$) sources at $z \geq 6$  with properties resemble those seen in strong LyC leakers, such as compact morphologies, steep UV slopes or high sSFRs \citep[e.g.,][]{morishita2020, bouwens2021, harikane2022}. However, the inferred number density of these sources is still largely uncertain, ranging from $\lesssim 10^{-7}$~Mpc$^{-3}$ to some $10^{-6}$~Mpc$^{-3}$ for $M_{\rm UV} \simeq -23$ at $z\simeq 7-8$ (e.g., \citealt{calvi2016, bowler2020, rojasruiz2020, finkelstein2022, leethochawalit2022}).

\subsection{Massive inflows feeding a UV-luminous starburst}\label{section45}

\begin{figure*}
  \centering
  \includegraphics[width=0.95\textwidth]{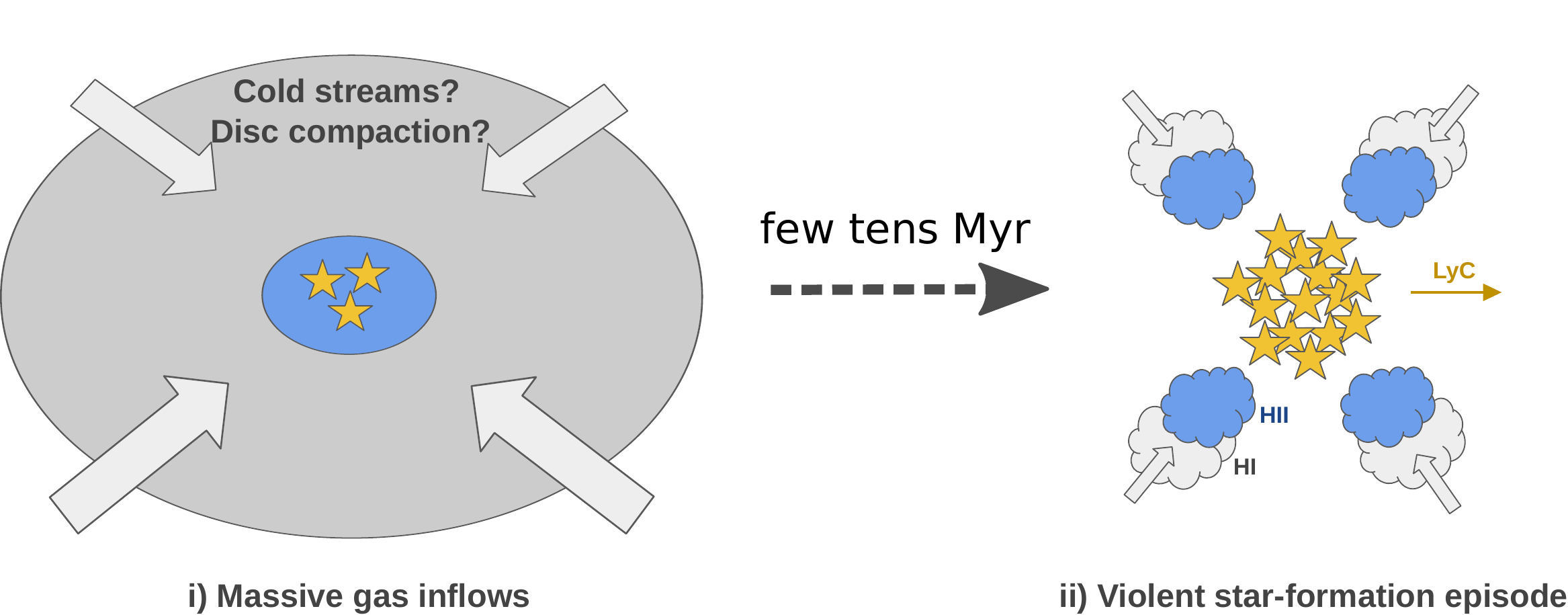}
  \vspace{5mm}
  \caption{Schematic temporal evolution of the proposed mechanism responsible for the massive starburst observed in J1316$+$2614. i) Disc contraction or bulk neutral gas inflows in J1316$+$2614 (left) would led to ii) an intense and concentrated star-formation in a short period of time of a few Myr (right), increasing the $sSFR$ and $\Sigma SFR$. The right panel also shows a schematic distribution and kinematics of the gas. LyC photons (yellow arrow) would escape from regions with low column density of neutral gas, $N_{\rm HI} \lesssim  10^{16}$~cm$^{-2}$. 
  }
  \label{fig9}
\end{figure*}

J1316$+$2614 is the most luminous star-forming galaxy known in the UV ($M_{\rm UV}=-24.68 \pm0.08$) and Ly$\alpha$ (log($L \rm [Ly\alpha / erg$~s$^{-1}] = 44.09\pm0.10$). We have shown that the large luminosity is powered by a young ($\simeq 10$~Myr) stellar population with $M_{\star}/M_{\odot} \simeq 10^{9.7}$ and $\rm SFR \simeq 500$~$M_{\odot}$~yr$^{-1}$ (Table~\ref{table4}), without any evidence of AGN activity. This yields a very high  $sSFR\simeq 105$~Gyr$^{-1}$ if only the mass of the young stellar population is assumed. Here we discuss possible mechanisms that could explain the intense starburst nature of J1316$+$2614.

With the available data we do not find any hint of a major merger that could enhance the SFR observed in J1316$+$2614. The morphology of J1316$+$2614 appears compact ($r_{\rm eff} \leq 2$kpc), and in addition, merger-induced SFRs are usually low, a factor of two at most (e.g., \citealt{pearson2019}). Rather than that, the rest-frame UV spectrum of J1316$+$2614 shows signs of inflowing gas and these are likely related to its recent SFH. 
The spectral shape of the Ly$\alpha$ emission in J1316$+$2614, with a blue peak more intense than the red one (Figure \ref{fig2}), and the redshifted ISM absorption of C~{\sc iv} (Figure \ref{fig1}) suggest gas inflows. 
Only a few star-forming galaxies are known to show similar Ly$\alpha$ profiles (e.g., \citealt{trainor2015}, \citealt{izotov2020}, \citealt{furtak2022}), but with much lower SNR. Redshifted ISM absorption lines have been also found in the spectra of star-forming galaxies at moderately low-$z$ (with a $\sim 5\%-6\%$ detection rates, e.g., \citealt{martin2012}, \citealt{rubin2012}), but are more elusive at high-$z$ (e.g., \citealt{falgarone2017}, \citealt{marques2018}, \citealt{herrera-camus2020}), possibly due to cosmological dimming and low SNR continuum spectra.

Cosmological simulations predict that massive inflows towards a central star-forming region from wet disc compaction may be frequent in high redshift galaxies \citep[][]{dekel2014, zolotov2015, dekel2020}. Violent disc instabilities, counter-rotating gas, gas-rich (minor) mergers, recycled gas inflows from galactic fountains, may provoke a dissipative shrinkage of gaseous discs, triggering the star formation in the central part of the galaxy, increasing the $sSFR$ and gas density. This event is usually referred to the "blue nugget" phase. At later times, this can eventually lead to central depletion and quenching of the star-formation \citep[e.g.,][]{tacchella2016}, resulting in a compact quiescent galaxy ("red nugget"). 

We speculate that a similar mechanism could be responsible for the massive starburst observed in J1316$+$2614. Figure~\ref{fig9} shows a schematic illustration of this mechanism that helps to explain and visualize several properties observed in J1316$+$2614. Gas contraction or neutral gas inflows in J1316$+$2614 (left) would lead to an intense and concentrated star-formation episode, increasing the $sSFR$ and $\Sigma SFR$ in a short period of time of a few Myr (right). 
This could explain the recent and bursty SFH, large luminosity, SFR, and the overall SED of J1316$+$2614 (right). The inflowing gas that we observed in J1316$+$2614 could correspond to gas left that has not been converted into stars. Gas in the remaining regions could have been consumed already in the starburst or removed by stellar winds and supernovae feedback. These regions would present low column density of neutral gas, $N_{\rm HI} \lesssim  10^{16}$~cm$^{-2}$, where LyC photons are escaping. 

Simulations predict that the maximum compaction of the gaseous disc occurs at stellar masses $\sim 10^{9.5}$~$M_{\odot}$ at $z\sim 2-4$  \citep{tacchella2016}, which is broadly consistent with that inferred for J1316$+$2614 ($M_{\star}/M_{\odot} \simeq 10^{9.67}$). 
Simulations also predict that the maximum increase of the $sSFR$ with respect to the MS is $\Delta \rm log(sSFR) \simeq 0.3-0.7$ in the compaction phase \citep{zolotov2015, tacchella2016b}, while for J1316$+$2614 we infer $\Delta \rm log(sSFR) \simeq 1.5$  if only the mass formed in the most recent star-formation history is considered. This suggests the presence of an old stellar population, that according to our results inferred in Section~\ref{section34} should be less massive than $M_{\star}^{\rm old}/M_{\odot} \leq 10^{10.3}$ (5$\sigma$). If both young and old stellar components are considered, the  specific SFR can decrease considerably, down to $sSFR \geq 20$~Gyr$^{-1}$ or $\geq 0.6$\,dex above the MS (Figure \ref{fig4}), but still consistent with simulations \citep[][]{zolotov2015, tacchella2016b}.

The violent star-formation episode observed in J1316$+$2614 resemble those of some massive and quiescent galaxies found at $z\gtrsim 3$, where strong ($SFR \gtrsim 1000$~$M_{\odot}$~yr$^{-1}$) and short ($\sim 50$~Myr) events of star formation are invoked \citep[e.g.,][]{glazebrook2017, forrest2020, valentino2020}.
Other works have identified compact star-forming galaxies with morphological properties and colors that are expected in the progenitors of quiescent and massive galaxies \citep[e.g., ][]{barro2013, barro2017, huertas2018}, but without the information on the gas kinematics. 
At some extent, similar mechanisms are required to explain the properties of the two star-forming galaxies known brighter than $M_{\rm UV} < -24.0$, BOSS-EUVLG1 at $z=2.47$ ($M_{\rm UV} = -24.4$; \citealt{marques2020b}) and J0121$+$0025 at $z=3.24$ ($M_{\rm UV} = -24.2$, also a LyC leaker; \citealt{marques2021}), both selected in the same way as J1316$+$2614 using the public BOSS/SDSS survey. These sources share similar properties as J1316$+$2614, characterized by young ($\lesssim 10$~Myr) starbursts with $\rm SFR \simeq 1000$~$M_{\odot}$~yr$^{-1}$, $sSFR \simeq 100$~Gyr$^{-1}$, and compact morphologies ($r_{\rm eff} \simeq 1$~kpc). 
Nevertheless, neutral and ionized outflows are detected in BOSS-EUVLG1 and J0121$+$0025  \citep{marques2020b, alvarez2021, marques2021}, without any evidence for the inflowing signatures that we observe in J1316$+$2614, but are also expected due to the increase of $sSFR$ and $\Sigma SFR$ in these sources \citep[e.g.,][]{williams2015}.

In short, J1316$+$2614 and the other UV-bright star-forming galaxies discovered recently \citep{marques2020b, marques2021} could represent the initial phases ($\sim 10$~Myr) in the evolution of massive galaxies. Their fate will depend on the amount of gas available and the efficiency to form new stars, which are still unknown. Star-formation can be quenched by starvation or they can continue to form new stars and dust, possibly reaching the far-IR bright phase. Other physical processes, e.g., major merger, formation of a super massive black hole or its ignition and feedback, could alter their fate as well. In any case, these UV-luminous phases must be shorted lived, lasting some tens to a few hundreds Myr, as already discussed in \cite{marques2020b}, but should mark drastic transitions on the properties of these galaxies, on the stellar mass build-up, chemical an dust enrichment, quenching mechanisms, and possibly on LyC leakage.

\section{Conclusion}\label{conclusion}

In this work we have presented the discovery and analysis of J1316$+$2614 at $z=3.613$, a luminous star-forming galaxy with high escape fraction of Lyman continuum radiation. While selected first as a bright QSO within BOSS/SDSS, follow-up observations with the GTC have revealed its true, starburst nature without any signs of AGN activity. From the analysis of these data we arrive at the following main results:

\begin{itemize}
    \item J1316$+$2614 is the most luminous star-forming galaxy known so far in the UV and Ly$\alpha$, with $M_{\rm UV}=-24.68 \pm0.08$ and log($L \rm [Ly\alpha / erg$~s$^{-1}] = 44.09\pm0.10$. The detection of stellar features in the rest-frame UV spectrum, such as photospheric absorption lines and wind lines, narrow nebular emission ($<500$~km~s$^{-1}$), and blue SED ($\beta_{\rm UV} = -2.59 \pm 0.05$, $r-W1 < -0.3$) discards a dominant AGN contribution to the luminosity of J1316$+$2614. We do not find any evidence of J1316$+$2614 being magnified by gravitational lensing. \vspace{2mm}

    \item The rest-frame UV spectrum is well reproduced by a S99 model with a continuous star-formation history with an age of 9~Myr, $Z_{\star}/Z_{\odot} \simeq 0.4$ and little dust obscuration ($E(B-V)=0.006^{+0.018}_{-0.006}$). The corresponding ionizing photon production efficiency is log($\xi_{\rm ion}[\rm Hz~erg^{-1}]) \approx 25.40$. 
    The optical to mid-IR photometry ($0.1-1.0\mu$m, rest) is dominated by the emission of this young stellar population. Our multi-wavelength best-fit SED model predicts a $SFR=497\pm92$~$M_{\odot}$~yr$^{-1}$ and log($M_{\star}/M_{\odot}) =9.67\pm0.07$ for the young stellar population. The presence of an old stellar population ($\geq 200$~Myr) is not well constrained, but should be less massive than log($M_{\star}/M_{\odot}) \leq 10.3$ (5$\sigma$). \vspace{2mm}

    \item LyC emission ($\lambda_{\rm 0} < 912$\AA) is significantly detected down to $830$\AA{ }and with a mean flux density $f_{\rm LyC} =1.69\pm0.10\mu$Jy. We infer a relative (absolute) LyC escape fraction $f_{\rm esc} \rm (LyC) \simeq 0.92$ ($\simeq 0.87$) assuming a relatively high IGM transmission ($\simeq 0.59$). The contribution of a foreground or AGN contamination to the LyC signal is unlikely. Other indirect tracers also suggest high escape fraction, including high specific SFR ($sSFR = 105 \pm 49$~Gyr$^{-1}$) and SFR surface density ($\Sigma \rm SFR \geq 10$~$M_{\odot}$~yr$^{-1}$~kpc$^{-2}$), and weak/non-detected low-ionization ISM lines. J1316$+$2614 is the most powerful ionizing source known among the star-forming galaxy population, both in terms of production ($Q_{\rm H} \approx 10^{56}$~s$^{-1}$) and escape ($f_{\rm esc} \rm (LyC) \approx 0.9$) of ionizing photons. 
    \vspace{2mm}
    
    \item Nebular emission is detected in Ly$\alpha$ and in rest-frame optical lines H$\beta$, [O~{\sc ii}] and [O~{\sc iii}], but these are much weaker ($EW_{0}$'s of Ly$\alpha$ and H$\beta$ of $\simeq 20-30$\AA) than that expected for the derived star-formation history of J1316$+$2614 ($EW \geq 120$\AA{ }for $\simeq 10$~Myr age). Our results demonstrate, for the first time, that large escape of ionizing photons affects strongly the strength of nebular lines and continuum emission, roughly by a factor of ($1- f_{\rm esc} \rm (LyC)$), and that only a fraction of ionizing photons will contribute to the photoionization of H~{\sc ii} regions. This suggests that at least some weak emission line galaxies could be potentially very strong LyC leakers. This may help in designing future surveys to detect very strong LyC emitters, which are for now restricted mostly in targeting extreme emission line star-forming galaxies, like green pea-like galaxies.  \vspace{2mm}
    
    \item The Ly$\alpha$ emission in J1316$+$2614 shows a double peak profile, with a blue peak more intense than the red one ($I_{\rm blue}/I_{\rm red}=3.7\pm0.1$) indicating neutral gas inflows. This is supported by the detection of redshifted ISM components of C~{\sc iv}. We speculate that massive inflows or dissipative compaction of the gas disc have triggered an intense and concentrated star-formation in the central part of the J1316$+$2614 and in a short period of time. This would explain its recent star-formation history, as well as many other properties observed in J1316$+$2614, such us its large luminosity, SFR, sSFR, $\Sigma \rm SFR$, SED, and gas kinematics. 
    J1316$+$2614 may represent the first case known of a star-forming galaxy undergoing a "blue nugget" phase, where gas inflows and a massive, compact and young starburst are observed simultaneously. 
    
\end{itemize}

\section*{Acknowledgements}

The authors thank the anonymous referee for useful comments that greatly improved the clarity of this work. R.M.C. thanks Fabrice Martins, Benjamin Magnelli and Jorge Sanchez-Almeida for useful discussions. 
Based on observations made with the Gran Telescopio Canarias (GTC)  installed in the Spanish Observatorio del Roque de los Muchachos of the Instituto de Astrof\'{i}sica de Canarias, in the island of La Palma. 
J.A.M., L.C., and I.P.F. acknowledge support from the Spanish State Research Agency (AEI) under grant numbers ESP2015-65597-C4-4-R, ESP2017-86852-C4-2-R, PID2021-127718NB-I00, PGC2018-094975-B-C22, and MDM-2017-0737 Unidad de Excelencia ''Mar\'{i}a de Maeztu''- Centro de Astrobiolog\'{i}a (CSIC-INTA). D.C. is a Ramon-Cajal Researcher and is supported by the Ministerio de Ciencia, Innovaci\'{o}n y Universidades (MICIU/FEDER) under research grant PID2021-122603NB-C21. A.S.L. acknowledge support from Swiss National Science Foundation. ASL acknowledge support from Swiss National Science Foundation. 

\section*{Data availability}
The data underlying this article will be shared on reasonable request to the corresponding author.

\bibliographystyle{mnras}
\bibliography{adssample_v2}

\label{lastpage}
\end{document}